\documentclass[twoside]{article}
\usepackage{amssymb}
\usepackage{amsmath}
\usepackage{amsthm}

\usepackage{qic,epsfig}  

\usepackage[usenames,dvipsnames]{pstricks}
\usepackage{pstricks}
\usepackage{url} 
\usepackage{graphicx}
\usepackage{enumerate}  

\newcommand*{\id}{{\bf 1}} 
 
\newcommand*{\cA}{\mathcal{A}}

\newcommand*{\cH}{\mathcal{H}}
\newcommand*{\cI}{\mathcal{I}}

\newcommand*{\cK}{\hat{\chi}}
\newcommand*{\cL}{\mathcal{L}}

\newcommand*{\cP}{\mathcal{P}}

\newcommand*{\cU}{\mathcal{U}}

\newcommand*{\ket}[1]{|#1\rangle}
\newcommand*{\bra}[1]{\langle #1|}
\newcommand*{\proj}[1]{\ket{#1}\bra{#1}}

\newcommand*{\maxt}{n} 
\newcommand*{\countsys}{I} 
\newcommand*{\counterv}{i}

\psset{unit=0.35cm}
\newcommand*{\zzero}{\psset{linecolor=gray}\qdisk(0,0){2.0pt}}
\newcommand*{\zzerobig}{\zzero}
\newcommand*{\zone}{\psset{linecolor=black}\qdisk(0,0){2.0pt}}
\newcommand*{\zundefined}{\psset{linecolor=red}\qdisk(0,0){2.0pt}}
\newcommand*{\zundefinedsecond}{\psset{linecolor=blue}\qdisk(0,0){2.0pt}}
\newcommand*{\zundefinedthird}{\psset{linecolor=green}\qdisk(0,0){2.0pt}}
\newcommand*{\zundefinedfourth}{\psset{linecolor=yellow}\qdisk(0,0){2.0pt}}

\newcommand*{\conjugation}{C_{\leftarrow}}
\newcommand*{\zdown}{\psset{linecolor=blue}
\psline{<-}(0,0)(1,1)
}

\newcommand*{\zup}{\psset{linecolor=blue}
\psline{<-}(0,0)(1,-1)
}

\newcommand*{\zleft}{\psset{linecolor=blue}
\psline{<-}(0,0)(1,0)
}

\newcommand*{\myhexagon}{\pspolygon[linewidth=0.7pt](0.5774;30)(0.5774;90)(0.5774;150)(0.5774;210)(0.5774;270)(0.5774;330)}

\newcommand*{\myhexagonthick}{\pspolygon[linewidth=2pt](0.5774;30)(0.5774;90)(0.5774;150)(0.5774;210)(0.5774;270)(0.5774;330)}

\newcommand*{\myhexagonsecond}{\pspolygon[linecolor=gray,linewidth=0.7pt](0.5774;30)(0.5774;90)(0.5774;150)(0.5774;210)(0.5774;270)(0.5774;330)}

\newcommand*{\myvertex}{\psline(0,0)(0.5774;90)\psline(0,0)(0.5774;330)\psline(0,0)(0.5774;210)}

\begin{document}
\setlength{\textheight}{8.0truein}    

\runninghead{Simplifying quantum double Hamiltonians using perturbative gadgets}
            {R. K\"onig}

\normalsize\textlineskip
\thispagestyle{empty}
\setcounter{page}{1}

\copyrightheading{10}{3 \& 4}{2010}{292--324}

\vspace*{0.88truein}

\alphfootnote

\fpage{1}

\centerline{\bf
SIMPLIFYING QUANTUM DOUBLE HAMILTONIANS}
\vspace*{0.035truein}
\centerline{\bf USING PERTURBATIVE GADGETS}
\vspace*{0.37truein}
\centerline{\footnotesize
ROBERT K\"ONIG}
\vspace*{0.015truein}
\centerline{\footnotesize\it Institute for Quantum Information, California Institute of Technology}
\baselineskip=10pt
\centerline{\footnotesize\it  Pasadena, California 91125, USA}
\publisher{May 13, 2009}{November 11, 2009}

\vspace*{0.21truein}

\abstracts{
Perturbative gadgets were originally introduced to generate effective $k$-local  interactions in the low-energy sector of a $2$-local Hamiltonian. Extending this idea, we present gadgets which are specifically suited for realizing Hamiltonians exhibiting non-abelian anyonic excitations. At the core of our construction is a perturbative analysis of a widely used hopping-term Hamiltonian. We show that in the low-energy limit,  this Hamiltonian can be approximated by a certain ordered product of operators. In particular, this provides a simplified realization of Kitaev's quantum double Hamiltonians.
}{}{}

\vspace*{10pt}

\vspace*{3pt}
\communicate{to be filled by the Editorial}

\vspace*{1pt}\textlineskip    

\section{Introduction}
The anyon based quantum computer, originally proposed by Kitaev~\cite{kitaev97,ogburnpreskill99,freedmanetal01}, is a promising approach to the realization of fault-tolerant quantum computing. Its main feature is the fact that quantum information is encoded in non-local observables related to topological invariants; such information is robust against local errors. The physical substrate of these schemes is a system with anyonic excitations above a degenerate ground space. Ideally, all computations are performed by successive creation, braiding and fusion of anyons, resulting in  non-trivial actions on the ground space.

A number of systems exhibiting the required topological ground state degeneracy and anyonic excitations are known: These include Kitaev's well-known $\mathbb{Z}_2$-toric code  and its generalization to arbitrary finite groups~\cite{kitaev97}, Kitaev's  model on the honeycomb lattice~\cite{kitaev06} and Levin and Wen's string-net models~\cite{LevinWen04}. The merits of a given proposal may be judged according to the following criteria: (i)~the difficulty of realizing the Hamiltonian in a physical system and (ii)~the computational power of its anyonic excitations. In terms of criterion~(i), the model~\cite{kitaev06} is perhaps most exciting, since the terms in the Hamiltonian are $2$-local interactions between neighboring qubits; indeed, corresponding experimental proposals already exist~\cite{duanetal03,michelietal06}. Unfortunately, the anyons in this model are not computationally universal. On the other hand, Levin and Wen's construction realizes all ``doubled'' topological phases, giving in particular anyons corresponding to  all discrete gauge theories and doubled Chern-Simons theories, but this model generally requires $12$-local interactions between qudits and therefore seems hard to realize. 

Perturbation theory is traditionally used to obtain an effective low-energy description of a given physical system. Perturbative gadgets turn this procedure around: they were introduced in~\cite{kempekitaevregev06} to realize certain ``target'' Hamiltonians as the low-energy limit of $2$-local Hamiltonians. This powerful idea has a number of applications~\cite{kempekitaevregev06,oliveiraterhal05,biamonte08,biamontelove08}; in~\cite{oliveiraterhal05}, it was observed that perturbative gadgets can generate (arbitrary) $k$-local effective Hamiltonians from $2$-local Hamiltonians. A particularly elegant construction achieving this was recently given by Jordan and Farhi~\cite{jordanfarhi} based on Bloch's perturbation series~\cite{bloch58}. While appealing in their generality and simplicity, these results may be of limited use for the realization of concrete Hamiltonians because of their large overhead: The number of required qubits and interactions scales polynomially in the number of non-trivial summands and factors when the target Hamiltonian is expressed in terms of Pauli operators. 

 It is natural to assume   that gadgets adapted to the internal structure of a given target Hamiltonian may lead to more efficient and natural realizations;  this is indeed the case for certain Hamiltonians that give rise to topological order. Here we focus on Kitaev's model~\cite{kitaev97} based on a finite group~$G$. Depending on the choice of~$G$, this has the following properties: if $G$ is any non-solvable group~ or a certain semidirect product of cyclic groups of prime order (which includes $G=S_3$), then the resulting anyons can be used to realize a universal gate set, as shown in a sequence of works~\cite{preskill97,ogburnpreskill99,mochon03,mochon04}.  The degrees of freedom in this model are $(d=|G|)$-qudits  placed on the edges of a lattice~$\cL$. The Hamiltonian has the form
\begin{align*}
H_{\textrm{QD}}&=-\sum_{v} A(v)-\sum_{p} B(p)\ ,
\end{align*}
where for every vertex~$v$, the vertex-term $A(v)$ acts on the incident edges of $v$, and the plaquette-term $B(p)$ acts on the boundary of plaquette~$p$ (We will define these operators in detail below). In particular, the terms in the Hamiltonian are at most $l=\max\{d_{\cL},d_{\cL'}\}$-local (more explicitly: they act on $(\mathbb{C}^{|G|})^{\otimes l}$). Here $d_{\cL}$ and $d_{\cL'}$ are the maximal degrees of vertices in the primal and dual lattice, respectively. Concretely, this gives interactions acting on $(\mathbb{C}^{|G|})^{\otimes 4}$ on a rectangular lattice, and interactions acting on $(\mathbb{C}^{|G|})^{\otimes 6}$ on a hexagonal lattice.

\section{Results and Techniques}
We summarize our results, and sketch the involved constructions. More details will be given below; here we merely give a high-level overview of our methods.
\subsection{Quantum double Hamiltonians as effective Hamiltonians} 
We construct
\begin{enumerate}[(a)]
\item\label{it:firstmainclaim}
a Hamiltonian $H_{\textrm{vertex}}$ which has the sum $-\sum_v A(v)$ of the vertex-terms of $H_{QD}$ as its low-energy limit, and consists of terms acting on $(\mathbb{C}^{|G|})^{\otimes 2}\otimes\mathbb{C}^{d_{\cL}}$. In a similar way, we can obtain a Hamiltonian $H_{\textrm{plaquette}}$ which generates $-\sum_p B(p)$ and is made of terms acting on $(\mathbb{C}^{|G|})^{\otimes 2}\otimes\mathbb{C}^{d_{\cL'}}$.
\item\label{it:secondmainclaim}
a Hamiltonian $H_{\textrm{full}}$ whose effective low-energy Hamiltonian completely reproduces $H_{QD}$. This construction consists of terms acting on  $(\mathbb{C}^{|G|})^{\otimes 2}\otimes(\mathbb{C}^{l})^{\otimes 3}$.
\end{enumerate}
To clarify the simple structure of these Hamiltonians, we schematically depict the operators in $H_{\textrm{vertex}}$ and $H_{\textrm{plaquette}}$  which generate a given vertex-/plaquette-operator:
\psset{unit=0.8cm}
\begin{align}
\begin{matrix}
\raisebox{-0.35cm}{\begin{pspicture}(-0.5577,-0.577)(0.5577,0.577)
\SpecialCoor
\psset{linewidth=2pt}
\rput(0,0){\myvertex}
\end{pspicture}} &\qquad &\longrightarrow&\qquad
\raisebox{-0.35cm}{\begin{pspicture}(-0.5577,-0.577)(0.5577,0.577)
\SpecialCoor\psset{linewidth=1pt}\psset{linecolor=gray}\rput(0,0){\myvertex}\psset{linecolor=black}\psset{linewidth=2pt}
\qdisk(0,0){2.5pt}\psline[linewidth=2pt]{-}(0,0)(0.5774;90)
\end{pspicture}}+
\raisebox{-0.35cm}{\begin{pspicture}(-0.5577,-0.577)(0.5577,0.577)
\SpecialCoor\psset{linewidth=1pt}\psset{linecolor=gray}\rput(0,0){\myvertex}\psset{linecolor=black}\psset{linewidth=2pt}
\qdisk(0,0){2.5pt}\psline[linewidth=2pt]{-}(0,0)(0.5774;330)
\end{pspicture}}+\raisebox{-0.32cm}{\begin{pspicture}(-0.5577,-0.577)(0.5577,0.577)
\SpecialCoor\psset{linewidth=1pt}\psset{linecolor=gray}\rput(0,0){\myvertex}\psset{linecolor=black}\psset{linewidth=2pt}
\qdisk(0,0){2.5pt}\psline[linewidth=2pt]{-}(0,0)(0.5774;210)
\end{pspicture}}\\
\raisebox{-0.4cm}{\begin{pspicture}(-0.5577,-0.577)(0.5577,0.577)
\SpecialCoor
\rput(0,0){\myhexagonthick}
\end{pspicture}} &\qquad &\longrightarrow&\qquad
\raisebox{-0.4cm}{\begin{pspicture}(-0.5577,-0.577)(0.5577,0.577)
\SpecialCoor
\rput(0,0){\myhexagonsecond}
\psline[linewidth=2pt]{-}(0.5774;30)(0.5774;90)
\qdisk(0,0){2.5pt}
\end{pspicture}} +
\raisebox{-0.4cm}{\begin{pspicture}(-0.5577,-0.577)(0.5577,0.577)
\SpecialCoor
\rput(0,0){\myhexagonsecond}
\psline[linewidth=2pt]{-}(0.5774;30)(0.5774;330)
\qdisk(0,0){2.5pt}
\end{pspicture}}+
\raisebox{-0.4cm}{\begin{pspicture}(-0.5577,-0.577)(0.5577,0.577)
\SpecialCoor
\rput(0,0){\myhexagonsecond}
\psline[linewidth=2pt]{-}(0.5774;330)(0.5774;270)
\qdisk(0,0){2.5pt}
\end{pspicture}}+
\raisebox{-0.4cm}{\begin{pspicture}(-0.5577,-0.577)(0.5577,0.577)
\SpecialCoor
\rput(0,0){\myhexagonsecond}
\psline[linewidth=2pt]{-}(0.5774;270)(0.5774;210)
\qdisk(0,0){2.5pt}
\end{pspicture}}+
\raisebox{-0.4cm}{\begin{pspicture}(-0.5577,-0.577)(0.5577,0.577)
\SpecialCoor
\rput(0,0){\myhexagonsecond}
\psline[linewidth=2pt]{-}(0.5774;210)(0.5774;150)
\qdisk(0,0){2.5pt}
\end{pspicture}}+
\raisebox{-0.4cm}{\begin{pspicture}(-0.5577,-0.577)(0.5577,0.577)
\SpecialCoor
\rput(0,0){\myhexagonsecond}
\psline[linewidth=2pt]{-}(0.5774;150)(0.5774;90)
\qdisk(0,0){2.5pt}
\end{pspicture}}
\end{matrix}\label{eq:visualizationfirst}
\end{align}
The operators on the right act  only on a single edge of the lattice and an auxiliary system located at the vertex/center of the plaquette, respectively. This auxiliary system is represented by a black circle; for vertices, it is  $\mathbb{C}^{|G|}\otimes \mathbb{C}^{3}$, and $\mathbb{C}^{|G|}\otimes \mathbb{C}^{6}$ for plaquettes.

The operators in $H_{\textrm{full}}$ are obtained by a similar substitution rule, with the difference that all auxiliary systems are of the same form $\mathbb{C}^{|G|}\otimes \mathbb{C}^{6}$, and the operators on the right also act on auxiliary systems associated with neighboring plaquettes and vertices, respectively:
\psset{unit=0.8cm}
\begin{align}
\begin{matrix}
\raisebox{-0.64cm}{\begin{pspicture}(-0.977,-0.977)(0.977,0.817)
\SpecialCoor
\rput(0.5774;30){\myhexagonsecond}
\rput(0.5774;270){\myhexagonsecond}
\rput(0.5774;150){\myhexagonsecond}
\psset{linecolor=black,linewidth=2pt}
\rput(0,0){\myvertex}
\end{pspicture}}&\qquad&\longrightarrow&\qquad
\raisebox{-0.64cm}{\begin{pspicture}(-0.977,-0.977)(0.977,0.817)
\SpecialCoor
\rput(0.5774;30){\myhexagonsecond}
\rput(0.5774;270){\myhexagonsecond}
\rput(0.5774;150){\myhexagonsecond}
\psset{linecolor=black,linewidth=2pt}
\qdisk(0,0){2.5pt}\psline(0,0)(0.5774;90)
\psset{linecolor=gray}
\qdisk(0.5774;30){2.5pt}\qdisk(0.5774;150){2.5pt}
\end{pspicture}}+
\raisebox{-0.64cm}{\begin{pspicture}(-0.977,-0.977)(0.977,0.817)
\SpecialCoor
\rput(0.5774;30){\myhexagonsecond}
\rput(0.5774;270){\myhexagonsecond}
\rput(0.5774;150){\myhexagonsecond}
\psset{linecolor=black,linewidth=2pt}
\qdisk(0,0){2.5pt}\psline(0,0)(0.5774;330)
\psset{linecolor=gray}
\qdisk(0.5774;30){2.5pt}\qdisk(0.5774;270){2.5pt}
\end{pspicture}}+
\raisebox{-0.64cm}{\begin{pspicture}(-0.977,-0.977)(0.977,0.817)
\SpecialCoor
\rput(0.5774;30){\myhexagonsecond}
\rput(0.5774;270){\myhexagonsecond}
\rput(0.5774;150){\myhexagonsecond}
\psset{linecolor=black,linewidth=2pt}
\qdisk(0,0){2.5pt}\psline(0,0)(0.5774;210)
\psset{linecolor=gray}
\qdisk(0.5774;150){2.5pt}\qdisk(0.5774;270){2.5pt}
\end{pspicture}}\\
\\
\raisebox{-0.4cm}{\begin{pspicture}(-0.5577,-0.577)(0.5577,0.577)
\SpecialCoor
\rput(0,0){\myhexagonthick}
\end{pspicture}} &\qquad&\longrightarrow&\qquad
\raisebox{-0.4cm}{\begin{pspicture}(-0.5577,-0.577)(0.5577,0.577)
\SpecialCoor
\rput(0,0){\myhexagonsecond}
\psline[linewidth=2pt]{-}(0.5774;30)(0.5774;90)
\qdisk(0,0){2.5pt}
\psset{linecolor=gray}
\qdisk(0.5774;30){2.5pt}\qdisk(0.5774;90){2.5pt}
\end{pspicture}} +
\raisebox{-0.4cm}{\begin{pspicture}(-0.5577,-0.577)(0.5577,0.577)
\SpecialCoor
\rput(0,0){\myhexagonsecond}
\psline[linewidth=2pt]{-}(0.5774;30)(0.5774;330)
\qdisk(0,0){2.5pt}
\psset{linecolor=gray}
\qdisk(0.5774;30){2.5pt}\qdisk(0.5774;330){2.5pt}
\end{pspicture}}+
\raisebox{-0.4cm}{\begin{pspicture}(-0.5577,-0.577)(0.5577,0.577)
\SpecialCoor
\rput(0,0){\myhexagonsecond}
\psline[linewidth=2pt]{-}(0.5774;330)(0.5774;270)
\qdisk(0,0){2.5pt}
\psset{linecolor=gray}
\qdisk(0.5774;330){2.5pt}\qdisk(0.5774;270){2.5pt}
\end{pspicture}}+
\raisebox{-0.4cm}{\begin{pspicture}(-0.5577,-0.577)(0.5577,0.577)
\SpecialCoor
\rput(0,0){\myhexagonsecond}
\psline[linewidth=2pt]{-}(0.5774;270)(0.5774;210)
\qdisk(0,0){2.5pt}
\psset{linecolor=gray}
\qdisk(0.5774;270){2.5pt}\qdisk(0.5774;210){2.5pt}
\end{pspicture}}+
\raisebox{-0.4cm}{\begin{pspicture}(-0.5577,-0.577)(0.5577,0.577)
\SpecialCoor
\rput(0,0){\myhexagonsecond}
\psline[linewidth=2pt]{-}(0.5774;210)(0.5774;150)
\qdisk(0,0){2.5pt}
\psset{linecolor=gray}
\qdisk(0.5774;210){2.5pt}\qdisk(0.5774;150){2.5pt}
\end{pspicture}}+
\raisebox{-0.4cm}{\begin{pspicture}(-0.5577,-0.577)(0.5577,0.577)
\SpecialCoor
\rput(0,0){\myhexagonsecond}
\psline[linewidth=2pt]{-}(0.5774;150)(0.5774;90)
\qdisk(0,0){2.5pt}
\psset{linecolor=gray}
\qdisk(0.5774;150){2.5pt}
\qdisk(0.5774;90){2.5pt}
\end{pspicture}}
\end{matrix}\label{eq:visualizationsecond}
\end{align}
Gray circles indicate that the action on these systems is particularly simple. (In fact, it is given by a projection onto a fixed vector $\ket{0}\in\mathbb{C}^{6}$ and leaves the first part $\mathbb{C}^{|G|}$ invariant.)

Comparing the support of the operators
in the original Hamiltonian $H_{QD}$ with those of the constructed Hamiltonians, we get the following table:
\vspace{0.1cm}
\begin{center}
\begin{tabular}{c|c||c|c}
lattice & $H_{QD}$ (original) &  $H_\textrm{plaquette}$, $H_\textrm{vertex}$  & $H_\textrm{full}$ \\
\hline
rectangular &  $(\mathbb{C}^{|G|})^{\otimes 4}$ & $(\mathbb{C}^{|G|})^{\otimes 2}\otimes\mathbb{C}^4$ &$(\mathbb{C}^{|G|})^{\otimes 2}\otimes(\mathbb{C}^4)^{\otimes 3}$\\
\hline
honeycomb  &  $(\mathbb{C}^{|G|})^{\otimes 6}$ & $(\mathbb{C}^{|G|})^{\otimes 2}\otimes\mathbb{C}^6$ &$(\mathbb{C}^{|G|})^{\otimes 2}\otimes(\mathbb{C}^6)^{\otimes 3}$
\end{tabular}
\end{center}
\vspace{0.2cm}
While the dimension of the support is decreased only for  $|G|>8$ on the rectangular lattice (and $|G|>3$ for the honeycomb lattice), we stress that the terms in our Hamiltonians are relatively simple: they involve at most $2$~systems related to the group~$G$ and are otherwise made of hopping-terms. The terms related to the group~$G$ are different versions of controlled-group multiplication, as explained below.

\subsubsection*{Obtaining the vertex-terms.} Let us sketch the construction of these Hamiltonians for the case of a square lattice. Consider a single vertex term~$A(v)$; this operator depends on the orientation of the edges of the lattice~$\cL$. For simplicity, assume that $v$ has edges $\{e_0^v,\ldots,e_3^v\}$ with arrows pointing away from $v$. In this case, the vertex operator is $A(v)=\frac{1}{|G|}\sum_g \prod_{i=0}^3 L^g_{-}(e_i^v)$, where the operator $L^g_{-}(e_i^v)$ stands for right-multiplication of the degree of freedom on edge $e_i^v$ by $g^{-1}$ for $g\in G$. Our aim is to find a Hamiltonian~$H^v$ whose effective low-energy Hamiltonian is equal to~$A(v)$.

We associate with the vertex~$v$ an auxiliary system $\cH_{R^v}\otimes\cH_{I^v}\cong \mathbb{C}^{|G|}\otimes\mathbb{C}^{4}$ with standard basis vectors $\ket{g}\ket{i}$, where $g\in G$ and $i=0,\ldots,3$. On $\cH_{R^v}$,
we introduce the projection operators
\begin{align*}
(T^g_+)_{R^v}&= \proj{g}_{R^v}\ ,\qquad g\in G\ .
\end{align*}
We then define the operators
\begin{align}
M_i^v &=\sum_{g} L^g_{-}(e_i^v)\otimes (T^g_+)_{R^v}\  \label{eq:mivdef}
\end{align}
acting on the degree of freedom associated with the edge $e_i^v$ and the auxiliary system $\cH_{R^v}$. Observe that this is a unitary. It corresponds to
controlled-right multiplication of the degree of freedom associated with the edge~$e_i^v$ by~$g^{-1}$, where~$g$ is the value of the auxiliary register. Our Hamiltonian is of the form  $H^v=H_0^v+\lambda V^v$ with
\begin{align}
H_0^v&=-\proj{\Psi^v}_{R^v}\otimes\proj{0}_{I^v}\ ,\nonumber\\
V^v &=\sum_{i=0}^3 \left((M_i^v)^\dagger \otimes\ket{i+1}\bra{i}_{I^v}+h.c.\right)\ , \label{eq:hzerov}
\end{align}
and $\ket{\Psi}_{R^v}=\frac{1}{\sqrt{|G|}}\sum_{g\in G}\ket{g}\in \cH_{R^v}$. (Here, h.c. denotes the Hermitian conjugate, and we omit identities when operators
act trivially on subsystems.) We show below that up to a constant energy shift, this Hamiltonian reproduces the term $-A(v)$ in $4$th order perturbation theory in~$\lambda$. 

To obtain the target Hamiltonian $-\sum_v A(v)$, we apply the described procedure
to every vertex~$v$. That is, we introduce an auxiliary system $\cH^v$ at every vertex, and define $H_0=\sum_v H_0^v$ and $V=\sum_v V^v$ as the sum over vertices of the operators defined by~\eqref{eq:hzerov}.  This  leads to claim~\eqref{it:firstmainclaim}.

\subsubsection*{Obtaining the plaquette-terms}
A similar statement holds for the plaquette-terms~$B(p)$. For concreteness, consider a plaquette whose boundary edges $\{e_0^p,\ldots,e_3^p\}$ are oriented in such a way that they form a 
counterclockwise loop around the plaquette. The plaquette-operator is equal to $B(p)=\sum_{g_3\cdots g_0=1}\prod_{i=0}^3 T^g_{-}(e^p_i)$, where the the sum is over all $4$-tuples $(g_0,\ldots,g_3)$ whose product $g_3g_2g_1g_0$ is equal to the identity element $1\in G$, and  $T^g_-(e^p_i)=\proj{g^{-1}}_{e^p_i}$ are projection operators acting on the degree of freedom on edge $e^p_i$. To generate this term in perturbation theory, we introduce auxiliary systems $\cH_{R^p}\otimes\cH_{I^p}\cong \mathbb{C}^{|G|}\otimes\mathbb{C}^4$ of the same form as before, and 
define the left-multiplication operator
\begin{align*}
(L^g_+)_{R^p} &=\sum_z \ket{gz}\bra{z}_{R^p}\ .
\end{align*}
on $\cH_{R^p}$. We then set
\begin{align}
M^p_i &=\sum_{g} T^g_-(e^p_i)\otimes (L^g_+)_{R^p} \label{eq:mpifirstdef}
\end{align}
and $\ket{\Psi^p}_{R^p}=\ket{1}\in\cH_{R^v}$.  The plaquette-term~$-B(p)$ is then again reproduced in $4$th order perturbation theory in $\lambda$ from $H^p=H^p_0+\lambda V^p$, with
\begin{align*}
H^p_0&= -\proj{\Psi^p}_{R^p}\otimes\proj{0}_{I^p}\\
V^p &= \sum_{i=0}^3 \left((M^p_i)^\dagger\otimes\ket{i+1}\bra{i}_{I^p}+h.c.\right)\ .
\end{align*}
The generalization to the target Hamiltonian $-\sum_p B(p)$ is straightforward.

\subsubsection*{Obtaining the full Hamiltonian $H_{QD}$.} 
The previously described naive procedure for generating several vertex- or plaquette-terms fails when we simultaneously consider vertex- and plaquette-terms. This is because the operators $M_i^v$ (cf.~\eqref{eq:mivdef}) associated with a vertex~$v$ generally do not commute with the operators $M_j^p$ (cf.~\eqref{eq:mpifirstdef}) associated with a plaquette~$p$ which has $v$ on its boundary. To overcome this problem, we introduce additional interactions: For any vertex~$v$, we modify the perturbation~\eqref{eq:hzerov} to
\begin{align*}
V^v &=\sum_{i=0} ^3\left( (M^v_i)^\dagger\otimes\ket{i+1}\bra{i}_{I^v}\otimes\proj{0}_{I^{p_{-}(e_i^v)}}\otimes\proj{0}_{I^{p_{+}(e_i^v)}}+h.c.\right)\ ,
\end{align*}
where $p_{-}(e_i^v)$ and $p_{+}(e_i^v)$ are the plaquettes separated by the edge $e_i^v$ (cf. Figure~\ref{fig:edgesimple} below); the expressions $V^p$ corresponding to plaquette-terms are changed in a similar way.  Our perturbative analysis then shows that the resulting perturbation 
$V=\sum_{v} V^v+\sum_p V^p$ generates the desired Hamiltonian $H_{QD}$ in the low-energy limit.

\subsection{Perturbative techniques for loop-Hamiltonians}
Our results rely on the following two facts. First, the vertex- and plaquette-terms of the quantum double Hamiltonian can be written as products of operators followed by a projection. Second, an ordered product of $k$~operators can be obtained in $k$-th order perturbation theory from a certain hopping-term Hamiltonian. 

The first fact boils down to the identities
\begin{align*}
A(v)&=\bra{\Psi^v}M_0^v\cdots M_3^v \ket{\Psi^v}\\
B(p)&=\bra{\Psi^p}M_0^p\cdots M_3^p \ket{\Psi^p}
\end{align*}
which can be verified by straightforward computation from the expressions given above.  However, more insight is gained from 
a derivation of this statement that uses some fundamental objects related to the quantum double Hamiltonian: We  first rewrite the vertex-term $A(v)$ as a ribbon-operator~\cite{kitaev97} associated with a closed ribbon (also called Wilson loop) going around the vertex~$v$. Operators corresponding to the concatenation of two ribbons may be written as a linear combination of products of ribbon-operators corresponding to their parts; this is described by the comultiplication map in the algebra of ribbon-operators. The operators~$(T^g_+)_{R^v}$ and $(L^g_+)_{R^p}$ used in the definition of $M_i^v$, $M_j^p$  are related to a representation of the dual Hopf algebra; these provide us with the desired expressions for $A(v)$ and $B(p)$ as  products. The procedure of writing the Hamiltonian in terms of operators corresponding to closed loops (ribbons), and subsequent decomposition of the loops into small segments could extend to similar quantum loop models.

The second fact may be of independent interest. We refer to it as the clock-gadget.
\subsubsection*{The Clock-gadget}
 This gadget provides a general tool for generating ordered products of operators. It is motivated by similar hopping-term constructions previously used to prove the QMA-completeness of the local Hamiltonian problem~\cite{kitaevetalbook02}. Roughly, we have the following statement.

\vspace*{12pt}
\noindent
\begin{theorem}(Informal version)
Let $H_0$ be a Hamiltonian on a bipartite system $\cH_S\otimes\cH_\countsys$ of the form $H_0=-P_0$, where $P_0=\Gamma_0\otimes\proj{0}_\countsys$ is the projection onto the ground space and where  $\{\ket{i}\}_{i=0}^{n-1}$ 
is an orthonormal basis of~$\cH_{I}$. Let 
\begin{align*}
V&=\sum_{\counterv=0}^{\maxt-1} \left(M_\counterv^\dagger \otimes \ket{\counterv+1}\bra{\counterv}+M_\counterv\otimes\ket{\counterv}\bra{\counterv+1}\right)\ ,
\end{align*} 
where we assume that the operators $\{M_\counterv\}_{\counterv}$ satisfy certain proportionality constraints.
Consider the Hamiltonian $H=H_0+\lambda V$.
 Up to a constant energy shift, the effective Hamiltonian (cf.~Section~\ref{sec:blochperturbation}) is
\begin{align}
H_{\textrm{eff}}&\approx (-1)^{n-1}\lambda^{\maxt}\left(\Gamma_0 M_0\cdots M_{\maxt-1}\Gamma_0+h.c.\right)+O(\lambda^{\maxt+1})\ .
\end{align}
\end{theorem}
\vspace*{12pt}
\noindent

The proof of this result is based on Bloch's formulation~\cite{bloch58} of degenerate perturbation theory and involves a certain diagrammatic formalism adapted to the problem at hand. The proportionality constraints in Theorem~\ref{thm:maingadget} hold in the case where the $M_i$'s are unitaries and are thus satisfied by the operators~\eqref{eq:mivdef} and~\eqref{eq:mpifirstdef}. Theorem~\ref{thm:maingadget} therefore gives a Hamiltonian which generates the vertex-term $A(v)$; the same procedure allows us to obtain a plaquette-operator~$B(p)$.
 \setcounter{theorem}{0}

We also provide a generalization of Theorem~\ref{thm:maingadget} to several clock systems~$\cH_{I^\alpha}$ with potentially non-commuting operators $\{M_i^\alpha,M_j^\beta\}$; this is needed to generate the full quantum double Hamiltonian $H_{QD}$.

\subsubsection*{Outline}
The remainder of the paper is structured as follows: In Section~\ref{sec:nonabelian}, we discuss the quantum double Hamiltonians and their representation in terms of loop operators. We then present the clock-gadget and its extensions in Section~\ref{sec:perturbation}. In Section~\ref{sec:perturbativegadgetsquantumdouble}, we combine these results to obtain our simplified Hamiltonians. We conclude in Section~\ref{sec:conclusions}.

\section{Non-abelian models based on the quantum double\label{sec:nonabelian}}
The goal of this section is to derive the necessary reformulation of the operators constituting the quantum double Hamiltonians defined by Kitaev~\cite{kitaev97}. We will first review these models (Section~\ref{sec:definitionhamiltonian})). We then give alternative expressions for the plaquette- and vertex-operators
as products~(Section~\ref{sec:plaquettevertexprod}). While the validity of this  reformulation can be checked immediately, its meaning is not immediately obvious, nor does it provide much insight into how our techniques may extend to similar models. In Section~\ref{sec:ribbonprodquantumdouble} we give what can be seen as a derivation of these expressions (This can be skipped at first reading). We relate plaquette- and vertex-operators to ribbon-operators, and show how representations of the quantum double allow us to factor the latter. This is the basis of our representation of the plaquette- and vertex-operators. Throughout, we mostly adopt Kitaev's notation~\cite{kitaev97}.

\subsection{Definition of the Hamiltonian\label{sec:definitionhamiltonian}}
 Kitaev's models~\cite{kitaev97} are based on the quantum double of a finite group~$G$. Consider a lattice $\cL$ on an orientable 2d-surface, and assign an orientation to each edge (represented by an arrow). We associate a quantum degree of freedom ($\cong \mathbb{C}^{|G|}$) with orthonormal basis states $\{\ket{g}_e\}_{g\in G}$ to every edge~$e$ of the lattice. The Hilbert space $\cH_{\cL}$ of the model is the tensor product of these spaces. We define the following operators (indexed by $g,h\in G$); they act on the degree of freedom associated with an edge $e$:
\begin{align*}
\begin{matrix}
L^{g}_{+}(e)&=\sum_{z}\ket{gz} \bra{z}_e\qquad & L^{g}_{-}(e)&=\sum_z \ket{zg^{-1}}\bra{z}_e\\
T^h_{+}(e) &= \proj{h}_e \qquad & T^h_{-}(e) &= \proj{h^{-1}}_e\ .
\end{matrix}
\end{align*}
Suppose that $e=(v_-,v_+)$, where  the arrow is directed from $v_-$ to $v_+$, and that $e$ separates  plaquettes $p_-$ (on the left) from plaquette $p_+$ (on the right, in the direction of the arrow). That is, $e$ has the form shown in Figure~\ref{fig:edgesimple}.   We  then set
\begin{align*}
\begin{matrix}
L^g(e,v_-)&=L^g_{-}(e)\ ,\qquad &L^g(e,v_+)&=L^g_{+}(e)\\
T^{h}(e,p_-)&=T^h_-(e)\ ,\qquad &T^h(e,p_+)&=T^h_+(e)\ .
\end{matrix}
\end{align*}
\begin{center}
\begin{figure}
\begin{center}
\begin{pspicture}(0,-0.5)(3,2.7)
\rput(1,-0.5){$v_{-}$}
\rput(0,1){$p_{-}$}
\rput(2,1){$p_{+}$}
\rput(1,2.5){$v_{+}$}
\psline{->}(1,0)(1,2)
\end{pspicture}
\end{center}
\fcaption{An edge $e$ with endpoints $v_-$ and $v_+$ separating two neighboring plaquettes $p_+$ and $p_-$.\label{fig:edgesimple}}
\end{figure}
\end{center}
The local gauge transformations $\{A_{g}(v,p)\}_{g\in G}$ and magnetic charge operators $\{B_g(v,p)\}_{g\in G}$ associated to a vertex~$v$ on the boundary of a plaquette~$p$ are
\begin{align}
\begin{matrix}
A_g(v,p) &=&A_g(v)=\prod_{e\in \textrm{star}(v)} L^g(e,v)\\
B_g(v,p) &=&\sum_{g_{k-1}\cdots g_0=g}\prod_{i=0} ^{k-1} T^{g_i}(\tilde{e}_i,p)\ ,
\end{matrix}\label{eq:gaugemagnetic}
\end{align}
where $\textrm{star}(v)$ denotes the set of edges incident to~$v$, and
where $\tilde{e}_0,\ldots,\tilde{e}_{k-1}$ are the edges on the  boundary of $p$~in clockwise order starting at the vertex~$v$. The Hamiltonian is expressed in terms of the vertex- and plaquette-operators
\begin{align*}
A(v)&=\frac{1}{|G|}\sum_{g\in G} A_g(v,p)\\
B(p)&=B_1(v,p)\ ,
\end{align*}
where $1\in G$ denotes the identity element; these are projections for every vertex~$v$ and plaquette~$p$. Note that $B(p)$ does not depend on which vertex~$v$ is chosen as the starting point. The  quantum double Hamiltonian on $\cH_{\cL}$ is defined as
\begin{align*}
H_{\textrm{QD}}&=-\sum_{v} A(v)-\sum_{p} B(p)\ .
\end{align*}

\subsection{Rewriting plaquette- and vertex-operators as products \label{sec:plaquettevertexprod}}
In this section, we present the main property of plaquette- and vertex-operators we need. We show that these operators can be written as ordered products of $2$-local operators acting on each edge of the plaquette/vertex and an auxiliary system, followed by a projection applied to the auxiliary system. 

More precisely, let $\cH_{R^p}\cong\mathbb{C}^{|G|}$  be an auxiliary system placed at the center of the plaquette~$p$. The operator $B(p)$ is  a product $M_0^p\cdots M_{n-1}^p$ of operators $M_i^p$ each acting on one edge~$e_i\in\partial p$ of the plaquette and the auxiliary system $\cH_{R^p}$; the auxiliary system has to be projected onto a certain state $\ket{\Psi^p}_{R^p}$. A similar statement holds for the vertex operator~$A(v)$. The support of the operators $\{M^v_i\}_i$ and $\{M^p_j\}_j$ is diagrammatically shown in~\eqref{eq:visualizationfirst}.

\vspace*{12pt} 
\noindent
\newtheorem{proposition}{Proposition}
\begin{proposition}\label{prop:specialrep}
Let $\{e_0,\ldots,e_{k-1}\}=\textrm{star}(v)$ be the edges incident to the vertex~$v$, enumerated in clockwise order starting from $e_0$. Similarly, let $\{\tilde{e}_0,\ldots,\tilde{e}_{j-1}\}=\partial p$ be the boundary edges of the plaquette~$p$ in clockwise order around $p$. 
Let $\cH_{R^v}\cong\cH_{R^p}\cong\mathbb{C}^{|G|}$ be auxiliary systems with orthonormal bases $\{\ket{g}\}_{g\in G}$, and let
\begin{align*}
T^g_{+}=\proj{g}_{R^v}\ \qquad\qquad L^g_{+}=\sum_{z}\ket{gz}\bra{z}_{R^p}
\end{align*}
 be projection onto~$\ket{g}$ and left-multiplication by~$g$ on these systems. Furthermore, let 
\begin{align*}
\ket{\Psi^v}_{R^v}= \frac{1}{\sqrt{|G|}}\sum_{g\in G}\ket{g}\qquad\textrm{and}\qquad
\ket{\Psi^p}_{R^p} = \ket{1}\ ,
\end{align*}
where $1\in G$ is the identity element.
Define the unitary operators
\begin{align*}
M_i^v=M(e_i,v) =\sum_{g} L^g(e_i,v)_{\cL}\otimes (T^g_+)_{R^v}\qquad\qquad M_i^p =M(\tilde{e}_i,p) =\sum_{g} T^g(\tilde{e}_i,p)_{\cL}\otimes (L^g_+)_{R^p}\ .
\end{align*}
on $\cH_{\cL}\otimes\cH_{R}$, which act non-trivially on one edge and the respective auxiliary system. Then
\begin{align*}
A(v)&= \bra{\Psi^v} M^v_0\cdots M^v_{k-1} \ket{\Psi^v}= \bra{\Psi^v} (M^v_{k-1})^\dagger \cdots (M^v_0)^\dagger \ket{\Psi^v}\\
B(p)&=\bra{\Psi^p} M^p_0\cdots M^p_{j-1}\ket{\Psi^p}=\bra{\Psi^p} (M^p_{j-1})^\dagger \cdots (M^p_0)^\dagger \ket{\tilde{\Psi^p}}\ .
\end{align*}
\end{proposition}
\vspace*{12pt} 
\noindent
{\bf Proof:} 
This  immediately  follows from the definitions. \square\,

To generate the complete quantum double Hamiltonian~$H_{QD}$ on $\cH_\cL$, we will later consider all plaquettes and vertices simultaneously. That is, we introduce auxiliary systems  $\cH_{R^v}\cong\cH_{R^p}\cong\mathbb{C}^{|G|}$  associated with each vertex~$v$ and plaquette~$p$, and consider the operators $\{M^v_i\}_{v,i}$ and $\{M^p_j\}_{p,j}$ defined by Proposition~\ref{prop:specialrep}. These act on the joint space $\cH_\cL\otimes(\bigotimes_v \cH_{R^v})\otimes(\bigotimes_p \cH_{R^p})$. For later reference, we show that they commute
unless they act on the same edge of~$\cL$, and are associated with a vertex and a plaquette, respectively.
\vspace*{12pt}
\noindent
\begin{lemma}\label{lem:miadditionalproperty}
Let $\{M_i^v,M_j^p\}$ be defined as above.
Then
\begin{enumerate}
\item $[M^v_i,M^{v'}_j]=[M^v_i,(M^{v'}_j)^\dagger]=0$ for all $v,v'$ and $i,j$. Similarly $[M^{p}_i,M^{p'}_j]=[M^{p}_i,(M^{p'}_j)^\dagger]=0$ for all~$p,p'$ and~$i,j$.\label{eq:commutmvjfirst}
\item 
$[M^v_i,M_j^p]=[M^v_i,(M_j^p)^\dagger]=0$ whenever these operators do not act on the same edge.
\label{eq:commutmvjsecond}
\item\label{eq:commutthird}
If $M_i^v=M(e,v_{\pm})$ and $M_j^p=M(e,p_{\pm})$ act on the same edge~$e$ (cf.~Figure~\ref{fig:edgesimple}), then $[M_i^v,M_j^p]\neq 0$.
\end{enumerate}
\end{lemma}
\vspace*{12pt}
\noindent
{\bf Proof:} 
To prove the first claim,  we can clearly restrict our attention to neighboring pairs of vertices $v_{-},v_{+}$ (or plaquettes $p_{-}$ and $p_{+}$) that are the endpoints of an edge $e=(v_{-},v_{+})$ (cf.~Figure~\ref{fig:edgesimple}). We can further assume that $M_i^{v_{-}}=M(e,v_{-})$ and $M_j^{v_{+}}=M(e,v_{+})$, i.e., that $M_i^{v_{-}}$ and $M_j^{v_{+}}$ both act on this edge ($e=e^{v_-}_i=e^{v_+}_j$). All other commutators vanish because the operators act on distinct systems. Observe that the action of $M^{v_-}_i$ on the degree of freedom associated with~$e$ is by right-multiplication by a group element, whereas $M_j^{v_{+}}$ acts by left-multiplication. The commutativity of these two operators therefore follows from the fact that left- and right-multiplication commute, i.e., $[L^g_-(e),L^h_+(e)]=0$ for all $g,h\in G$. This gives the first part of claim~\ref{eq:commutmvjfirst}. The second part of claim~\ref{eq:commutmvjfirst} is  derived in a similar fashion, using $[T^g_-(e),T^h_+(e)]=0$ for all $g,h\in G$.

Claim~\ref{eq:commutmvjsecond} immediately follows  from the definitions since the corresponding operators act on distinct systems. Finally, claim~\ref{eq:commutthird} can be verified by a straightforward computation using the definition of these operators. \square\,

In the next section, we connect the expressions given in Proposition~\ref{prop:specialrep} to ribbon-operators which create localized excitations. While this is not essential for our main result, we hope that this additional detail might provide some guide as to how to generalize to other models.

\subsection{Ribbon-operators,  the quantum double and  products\label{sec:ribbonprodquantumdouble}}
\begin{figure}
\psset{unit=1.2cm}
\begin{center}
\begin{pspicture}(-1.3,-1.7)(2.8,0.9)
\SpecialCoor
\pspolygon[linewidth=0.7pt,fillcolor=lightgray,fillstyle=solid](0,0)(1,0)(0.5,-0.2887)
\pspolygon[linewidth=0.7pt,fillcolor=gray,fillstyle=solid](1.5,-0.866)(1.5,-0.2887)(2,-0.5774)
\rput(0,0){\myhexagon}\rput(1,0){\myhexagon}\rput(2,0){\myhexagon}
\rput(-0.5,-0.866){\myhexagon}\rput(0.5,-0.866){\myhexagon}\rput(1.5,-0.866){\myhexagon}
\rput(0.5,-0.45){$v$}
\rput(1.4,-0.95){$p$}
\rput(0.6,0.1){$e$}
\rput(1.85,-0.26){$\tilde{e}$}
\end{pspicture}
\end{center}
\fcaption{The two types of triangles corresponding to the shortest possible ribbons:  the first one is described by an edge~$\tilde{e}$ and a plaquette~$p$; it will appear in the decomposition of $B(p)$. The second one is specified by a pair $(e,v)$, and will be used to decompose the vertex operator~$A(v)$.\label{fig:simpletriangle}}
\end{figure}
We first recall the definition and some basic properties of ribbon-operators following~\cite{kitaev97}. We will subsequently use a representation of the quantum double to rewrite plaquette- and vertex-operators in the desired form.

Quasiparticles in the quantum double model are associated with pairs $(v,p)$, where $v$ is a vertex on the boundary of plaquette $p$ (these are called sites).  A ribbon~$t$ connects two sites $(v,p)$ and $(v',p')$, and is defined by a path on~$\cL$ connecting $v$ and $v'$, and an associated path in the dual lattice~$\cL'$ connecting the center of~$p$ with that of~$p'$. For every ribbon $t$, there are $|G|^2$~ribbon-operators $\{F^{(h,g)}(t)\}_{(h,g)\in G^2}$ which create particle pairs at the ends of the ribbon. These operators generate an algebra by
\begin{align}
F^{\bf m}(t)F^{\bf n}(t)=\sum_{\bf k}\Lambda^{\bf mn}_{\bf k}F^{\bf k}(t)\qquad\textrm{ where }\qquad \Lambda^{(h_0,g_0),(h_1,g_1)}_{(h,g)}=\delta_{h_0h_1,h}\delta_{g_0,g}\delta_{g_1,g}\ .
\label{eq:ribbonmultiplic}
\end{align}
Ribbon-operators corresponding to a ``long'' ribbon $t=t_0t_1$ can be written as a linear combination of ribbon-operators corresponding to the ribbons $t_0$ and $t_1$ by
\begin{align}
F^{\bf k}(t_0t_1) &=\sum_{\bf m,n}\Omega^{\bf k}_{\bf mn} F^{\bf m}(t_0)F^{\bf n}(t_1)\ ,\label{eq:ribbonextension}
\end{align}
where
\begin{align}
\Omega^{(h,g)}_{(h_0,g_0),(h_1,g_1)}&=\delta_{g,g_0g_1}\delta_{h_0,h}\delta_{h_1,g_0^{-1}hg_0}\label{eq:omegatensordef}
\end{align}
(Mathematically,~\eqref{eq:ribbonextension} represents comultiplication in the Hopf algebra generated by the ribbon-operators, cf.~\cite{kitaev97}.)
The ribbon-operators corresponding to the shortest possible ribbons are those associated with a single triangle. These come in two types, see~Figure~\ref{fig:simpletriangle}. The first type of triangle has two sides pointing to the center of a plaquette~$p$ and an edge~$\tilde{e}$ on the boundary of~$p$ as its third side (we simply write $(\tilde{e},p)$ instead of $t$); the second type has two sides pointing to a vertex~$v$ and one side which is an edge $e'$ of the dual lattice $\cL'$. This type is specified by the edge $e$ crossed by $e'$ and its endpoint $v$. The associated operators are
\begin{align}\label{eq:simpleribb}
 F^{(h,g)}(\tilde{e},p)=T^{g^{-1}}(\tilde{e},p)\qquad\qquad F^{(h,g)}(e,v)=\delta_{g,1}L^h(e,v)\ .
\end{align}
Observe that~\eqref{eq:ribbonextension} and~\eqref{eq:simpleribb} completely determine the ribbon-operators.

We are interested in the operators associated to closed ribbons. More precisely, we consider closed ribbons around vertices of the primal lattice, and similarly for the dual lattice (see Figure~\ref{fig:closedribbon}). Let us define $t[v,e_0]$ to be the closed ribbon going around vertex $v$ starting from the edge~$e_0\in \textrm{star}(v)$ in a clockwise fashion. Similarly, let $t[p,\tilde{e}_0]$ be the closed ribbon going around the center of~$p$ starting from the edge $\tilde{e}_0$ in a clockwise fashion; notice that these are products of only one type of ``simple'' operators. In particular, if $v$ has incident vertices $e_0,\ldots, e_{j-1}\in\textrm{star}(v)$ (going clockwise starting from $e_0$),
and plaquette $p$ is surrounded by $\tilde{e}_0,\ldots,\tilde{e}_{k-1}$ as explained in Section~\ref{sec:nonabelian},
these ribbons can be decomposed into a sequence of triangles as
\begin{align}
t[v,e_0] =(e_0,v)(e_1,v)\cdots (e_{j-1},v)\qquad t[p,\tilde{e}_0]=(\tilde{e}_0,p)(\tilde{e}_1,p)\cdots (\tilde{e}_{k-1},p)\ .\label{eq:loopribbons}
\end{align}
When the choice of the starting edge $e_0$ (or $\tilde{e}_0$) is irrelevant, we will drop this in our notation  and simply write $t_v=t[v,e_0]$ and $t_p=t[p,\tilde{e}_0]$.
A straightforward computation then shows the following: 
\begin{figure}
\psset{unit=1.2cm}
\begin{center}
\begin{pspicture}(-1.3,-1.7)(2.8,0.9)
\SpecialCoor
\pspolygon[linewidth=0.7pt,fillcolor=lightgray,fillstyle=solid](0,0)(1,0)(0.5,-0.2887)
\pspolygon[linewidth=0.7pt,fillcolor=lightgray,fillstyle=solid](1,0)(0.5,-0.2887)(0.5,-0.866)
\pspolygon[linewidth=0.7pt,fillcolor=lightgray,fillstyle=solid](0,0)(1,0)(0.5,-0.866)
\rput(1.5,-0.866){\pspolygon[linewidth=0.7pt,fillcolor=gray,fillstyle=solid](0.5774;30)(0.5774;90)(0.5774;150)(0.5774;210)(0.5774;270)(0.5774;330)}
\rput(0,0){\myhexagon}\rput(1,0){\myhexagon}\rput(2,0){\myhexagon}
\rput(-0.5,-0.866){\myhexagon}\rput(0.5,-0.866){\myhexagon}\rput(1.5,-0.866){\myhexagon}
\rput(0.5,-0.45){$v$}
\rput(1.4,-0.95){$p$}
\end{pspicture}
\end{center}
\fcaption{In Lemma~\ref{lem:basicplaquettestar}, we show how the vertex-operator can be written as a sum of ribbon-operators corresponding to a ribbon going around vertex~$v$. An analogous statement holds for the plaquette-operator~$B(p)$. This figure schematically shows these ribbons.\label{fig:closedribbon}}
\end{figure}
\vspace*{12pt}
\noindent
\begin{lemma}\label{lem:basicplaquettestar}
The local gauge transformations $\{A_{g}(v,p)\}_{g\in G}$ and magnetic charge operators $\{B_g(v,p)\}_{g\in G}$ are ribbon-operators corresponding to closed loops, that is,
\begin{align*}
A_g(v,p) &= A_g(v)=F^{(g,1)}(t_v)\\
B_g(v,p) &=F^{(h,g^{-1})}(t[p,\tilde{e}_0])\qquad\textrm{ for all }h\in G\
 .
\end{align*}
In particular, the vertex- and plaquette-operators can be represented as
\begin{align*}
A(v) &=\frac{1}{|G|}\sum_{\bf k} \mathsf{e}^{\bf k} F^{\bf k}(t_v)\\
B(v,p)&=\frac{1}{|G|}\sum_{\bf k} \mathsf{e}^{\bf k} F^{\bf k}(t_p) =F^{(h,1)}(t_p)\qquad\textrm{ for all }h\in G\ ,
\end{align*}
where
\begin{align}
\mathsf{e}^{(h,g)}&=\delta_{g,1}\qquad\textrm{for all }(h,g)\in G^2\  .\label{eq:edeltadef}
\end{align}

\end{lemma} 
\vspace*{12pt}
\noindent
{\bf Proof:}  
 Consider a ribbon $t=t_0t_1$.  By~\eqref{eq:ribbonextension}, we have
\begin{align*}
F^{(h,g)}(t_0t_1) &=\sum_{g_0g_1=g}F^{(h,g_0)}(t_0)F^{(g_0^{-1}hg_0,g_1)}(t_1)\ .
\end{align*}
Consider the case of the local gauge transformations. We show inductively that for every $i=1,\ldots,j-1$
\begin{align}
F^{(h,g)}(t_0\cdots t_i) &=\delta_{g,1}L^{h}(e_0,v)\cdots L^h(e_i,v)\ , \label{eq:induction}
\end{align}
where $t_i=(e_i,v)$.
For the base case, we get from~\eqref{eq:simpleribb}
\begin{align*}
F^{(h,g)}(t_0t_1)&=\sum_{g_0g_1=g} \delta_{g_0,1}L^{h}(e_0,v)\delta_{g_1,1}L^{g_0^{-1}hg_0}(e_1,v)\\
&=\delta_{g,1}L^{h}(e_0,v)L^h(e_1,v)\ .
\end{align*}
Assume that~\eqref{eq:induction} holds for $i$. Then
\begin{align*}
F^{(h,g)}(t_0\cdots t_i t_{i+1}) &=\sum_{g_0g_1=g}F^{(h,g_0)}(t_0\cdots t_i) F^{(g_0^{-1}hg_0,g_1)}(t_{i+1})\\
&=\sum_{g_0g_1=g} \delta_{g_0,1}\left(\prod_{r=1}^i L^h(e_r,v)\right)\delta_{g_1,1}L^{(g_0^{-1}hg_0,g_1)}(e_{i+1},v)\ ,
\end{align*}
where we used the induction hypothesis and~\eqref{eq:simpleribb}. This immediately implies~\eqref{eq:induction} for all~$i$. This concludes the proof of the statement about the local gauge transformations and the vertex-operators~$A(v)$.

The other claims can be proved similarly, with~\eqref{eq:induction} replaced by
\begin{align*}
F^{(h,g)}(\tilde{t}_0\cdots \tilde{t}_i)&=\sum_{g_i\cdots g_0=g}T^{g_0}(\tilde{e}_0,p)\cdots T^{g_i}(\tilde{e}_i,p)\qquad\textrm{ for all }h\in G\ ,
\end{align*}
where $\tilde{t}_i=(\tilde{e}_i,p)$. \square\,

Our next goal is to write  the plaquette- and vertex-operators as products of operators. 
 We  use Drinfeld's quantum double~\cite{drinfeld87} $\mathcal{D}=\mathcal{D}(G)$ of the group~$G$.
This is a Hopf algebra with generators $\{D_{(h,g)}\}_{(h,g)\in G^2}$ satisfying the multiplication rule
\begin{align}
D_{\bf m}D_{\bf n}=\sum_{\bf k}\Omega^{\bf k}_{\bf mn} D_{\bf k}\ .\label{eq:quantumdouble}
\end{align}
Note that $\mathcal{D}$ is the dual of the algebra generated by the ribbon-operators. Note also that the local
gauge transformations and magnetic charge operators~\eqref{eq:gaugemagnetic} form a representation of $\mathcal{D}$ by $D_{(h,g)}=B_h(v,p) A_g(v,p)$. The algebra~$\mathcal{D}$ gives a way of writing~\eqref{eq:ribbonextension} as a product of operators. That is, suppose we have a representation of the algebra~$\mathcal{D}$. We then have the identity
\begin{align}\left(\sum_{\bf m} F^{\bf m}(t_0)\otimes D_{\bf m}\right)\left(\sum_{\bf n} F^{\bf n}(t_1)\otimes D_{\bf n}\right)&=\sum_{\bf m,n} F^{\bf m}(t_0)F^{\bf n}(t_1)\otimes D_{\bf m}D_{\bf n}\nonumber\\
&\underset{=}{\tiny\textrm{\eqref{eq:quantumdouble}}}\sum_{\bf m,n,k} \Omega^{\bf k}_{\bf mn}F^{\bf m}(t_0)F^{\bf n}(t_1)\otimes D_{\bf k}\nonumber\\
&\underset{=}{\tiny\textrm{\eqref{eq:ribbonextension}}}\sum_{\bf k} F^{\bf k}(t_0t_1)\otimes D_{\bf k}\label{eq:ribboncombined}\ \end{align}
In particular, by induction, we obtain the formula
\begin{align}
M_0\cdots M_{n-1} &=\sum_{\bf k}F^{\bf k}(t)\otimes D_{\bf k}\ ,\label{eq:productofms}
\end{align}
for a ribbon $t=t_0\cdots t_{n-1}$, where
\begin{align}
M_i&=M(t_i)=\sum_{\bf n} F^{\bf n}(t_i)\otimes D_{\bf n}\ .\label{eq:midef}
\end{align}
Identity~\eqref{eq:productofms} is our starting point for writing ribbon-operators as products. Because of Lemma~\ref{lem:basicplaquettestar}, we are especially interested in certain sums of ribbon-operators corresponding to closed ribbons. Such a sum can be obtained from~\eqref{eq:midef} by choosing
a particular representation of~$\mathcal{D}$, and projecting the second system onto a certain vector.

Concretely, we represent the algebra $\mathcal{D}$ on a Hilbert space $\cH_R\cong (\mathbb{C}^{|G|})^{\otimes 2}$ with orthonormal basis $\{\ket{{\bf k}}=\ket{k_0}\ket{k_1}\}_{{\bf k}\in G^2}$.  The action of the generating operators on these vectors is given by
\begin{align}
D_{\bf j} \ket{{\bf k}} &=\sum_{\bf m}\Omega^{\bf k}_{\bf mj}\ket{{\bf  m}}\ .\label{eq:drep}
\end{align}
The fact that this is a representation of the algebra~$\mathcal{D}$ follows from
the identity (cf.~\cite[(31)]{kitaev97})
\begin{align}
\sum_{\bf m}\mathsf{e}^{\bf m}\Omega^{\bf k}_{\bf mj}=\sum_{\bf m}\Omega^{\bf k}_{\bf jm}\mathsf{e}^{\bf m}=\delta^{\bf k}_{\bf j}\label{eq:kitsum}
\end{align}
 We define the normalized vector
\begin{align}
\ket{\Psi}_R &=\frac{1}{\sqrt{|G|}}\sum_{{\bf k}} \mathsf{e}^{{\bf k}}\ket{{\bf  k}}=
\frac{1}{\sqrt{|G|}}\sum_{h}\ket{(h,1)}\ .\label{eq:specialvector}
\end{align}
Focusing on the case where the ribbon goes around a vertex of the primal/dual lattice, we get the following statement.

\vspace*{12pt}
\noindent
\begin{lemma}\label{lem:specialrepsecond}
Let $t=t_0\cdots t_{n-1}$ be of the form $t_v$ or $t_p$~\eqref{eq:loopribbons}, and let $M_i=M(t_i)$ be defined by~\eqref{eq:midef}. Let $\ket{\Psi}$ be given by~\eqref{eq:specialvector}. Then
\begin{align*}
\bra{\Psi} M_0\cdots M_{n-1}\ket{\Psi}&=\bra{\Psi} M_{n-1}^\dagger \cdots M_0^\dagger \ket{\Psi}=\begin{cases}
A(v)\ \textrm{ if }t=t_v\textrm{ surrounds vertex }v\\
B(p)\ \textrm{ if }t=t_p\textrm{ surrounds plaquette }p\ .\\
\end{cases}\end{align*}
\end{lemma}
\vspace*{12pt}
\noindent
{\bf Proof:} Note that the coefficients $\mathsf{e}^{\bf k}$ are real.
Therefore
\begin{align}
\bra{\Psi} D_{\bf j}\ket{\Psi } &=\frac{1}{|G|}\sum_{{\bf m},{\bf k}} \mathsf{e}^{\bf m}\mathsf{e}^{\bf k} \Omega^{\bf k}_{\bf mj}
\end{align}
With~\eqref{eq:kitsum}, 
 this reduces to
\begin{align}\label{eq:Dimport}
\bra{\Psi} D_{\bf j}\ket{\Psi } &=\frac{1}{|G|}\sum_{{\bf k}} \mathsf{e}^{\bf k}\delta^{\bf k}_{\bf j}=\frac{1}{|G|}\mathsf{e}^{\bf j}\ .
\end{align}
Applying this identity to~\eqref{eq:productofms} using linearity gives
\begin{align*}
\bra{\Psi} M_0\cdots M_{n-1}\ket{\Psi}=\begin{cases}
A(v)\qquad\textrm{ if }t\textrm{ surrounds vertex }v\\
B(p)\qquad\textrm{ if }t\textrm{ surrounds center of plaquette }p\ 
\end{cases}
\end{align*}
because of Lemma~\ref{lem:basicplaquettestar}. Since the rhs.~of this equation is Hermitian, the claim follows by taking the adjoint. \square\,

To relate Lemma~\ref{lem:specialrepsecond} to the analogous statement of Proposition~\ref{prop:specialrep}, we derive more explicit expressions for the operators $M_i=M(t_i)$ defined by~\eqref{eq:midef}. 
For this purpose, consider the action of the operators $D_{(h,g)}$. We have
\begin{align*}
D_{(h_1,g_1)}\ket{(h,g)}&=\sum_{(h_0,g_0)}\Omega^{(h,g)}_{(h_0,g_0),(h_1,g_1)}\ket{(h_0,g_0)}\\
&=\sum_{(h_0,g_0)} \delta_{g_0,gg_1^{-1}}\delta_{h_0,h}\delta_{h_1,g_0^{-1} hg_0}\ket{(h_0,g_0)}\\
&=\delta_{h_1,(gg_1^{-1})^{-1}h(gg_1^{-1})}\ket{(h,gg_1^{-1})}\ .
\end{align*} 
From this, it is clear that
\begin{align*}
D_{(h_1,g_1)}&=\conjugation^{-1}(\proj{h_1}\otimes\id_{\mathbb{C}^{|G|}})\conjugation (\id_{\mathbb{C}^{|G|}}\otimes L^{g_1}_{-})\ ,
\end{align*}
where $\conjugation$ is the unitary operator defined by
\begin{align}
\conjugation\ket{(h,g)}&=\ket{(g^{-1} h g,g)}\qquad\textrm{for all }(h,g)\label{eq:conjugationdef}\ ,
\end{align}
and where $L^g_{-}$ denotes right-multiplication by $g^{-1}$.
For example, if $t_i=(e_i,v)$ is a triangle in a ribbon going around the vertex~$v$, we get 
\begin{align}
M(e_i,v) &=\sum_h L^h(e,v)\otimes \conjugation^{-1}(\proj{h}\otimes \id_{\mathbb{C}^{|G|}})\conjugation\ \label{eq:vertexm}
\end{align}
by~\eqref{eq:simpleribb}. Similarly, if  $t_i=(e_i,p)$ is associated with a triangle which is part of a ribbon going around the center of plaquette~$p$, we have
\begin{align}
M(e_i,p) &=\sum_g T^{g^{-1}}(e,p)\otimes\id_{\mathbb{C}^{|G|}}\otimes L^{g}_{-}\ .\label{eq:meipdef}
\end{align}
Observe that the operators~\eqref{eq:vertexm} and~\eqref{eq:meipdef} only act non-trivially on one factor of the auxiliary system~$\cH_R\cong (\mathbb{C}^{|G|})^{\otimes 2}$. With~\eqref{eq:specialvector},~\eqref{eq:vertexm} and~\eqref{eq:meipdef},  the expressions given in Lemma~\ref{lem:specialrepsecond} immediately reduce to the simpler expressions given in Proposition~\ref{prop:specialrep}.

\section{Perturbation theory for effective Hamiltonians\label{sec:perturbation}}
The outline of this section is as follows: In Section~\ref{sec:blochperturbation}, we give the relevant expressions of Bloch's perturbation expansion of effective Hamiltonians, adapted to the special case of interest for simplicity (see~\cite{jordanfarhi} for more details -- we closely follow this reference). In Section~\ref{sec:perturbativegadgetsordered}, we introduce our main tool, the clock-gadget. We then present an important generalization in Section~\ref{sec:multiclock}.

\subsection{Bloch's perturbation series\label{sec:blochperturbation}}
Let $H_0$ be a Hamiltonian, and let  $P_0=\sum_{i=1}^d \proj{\varphi^{(0)}_i}$
be the projection onto the $d$-fold degenerate ground space. Let $\{E^{(0)}_j\}_{j\geq 0}$ be the (ordered) eigenvalues of $H_0$. We assume that the ground state energy of $H_0$ is $E^{(0)}_0=0$. (If this is not the case, we simply consider the shifted Hamiltonian $H'_0=H_0-E^{(0)}_0\cdot \id$ since this will not affect our results.) Consider the perturbed Hamiltonian $H=H_0+\lambda V$; let $\{\ket{\varphi_i}\}_{i=1}^d$ be mutually orthogonal eigenstates of~$H$ corresponding to the $d$~lowest energies, and let $\{E_i\}_i$~be their energies. The effective Hamiltonian $H_{\textrm{eff}}$ for the $d$ lowest energy levels is defined as 
\begin{align*}
H_{\textrm{eff}}=\sum_{i=1}^d E_i\proj{\varphi_i}\ .
\end{align*}
The perturbation $\lambda V$ generally leads to a shift of the energy of the entire space; we are usually not interested in this aspect, but only in how the Hamiltonian acts on the span of the states $\ket{\varphi_i}$. Therefore, it is useful to introduce the shifted effective Hamiltonian
\[
\tilde{H}_{\textrm{eff}}(\delta) =H_{\textrm{eff}}-\delta\cdot \Pi\  ,
\]
where $\Pi=\sum_{i=1}^d \proj{\varphi_i}$ is the projection onto the support of $H_{\textrm{eff}}$, and $\delta$ parametrizes the magnitude of the energy shift.

Bloch's perturbation expansion gives an expression of the effective Hamiltonian as
\begin{align}
H_{\textrm{eff}}&=\cU\cA\cU^\dagger\ ,\label{eq:heffbloch}
\end{align}
with associated power series expansions of $\cA$ and $\cU$ in $\lambda$:\begin{align}
\cA &=\sum_{m=0}^{\infty}\lambda^m \cA^{(m)}\qquad\cU =\sum_{m=0}^{\infty}\lambda^m \cU^{(m)}\ .\label{eq:powerseriesexpansions}
\end{align}
The operator $\cU$ satisfies
\begin{align}
\Pi=\cU P_0\cU^\dagger\ .\label{eq:pzeroconstantshift}
\end{align}
The $m$-th terms in the power series expansions  are 
\begin{align}
\cA^{(0)}&=0\qquad\cU^{(0)}=P_0\label{eq:auzeroorder}
\end{align}
and for $m>0$
\begin{align}
\cA^{(m)}&=\sum_{(\ell_1,\ldots,\ell_{m-1})\in\cP_{m-1}} P_0 V S^{\ell_1}V S^{\ell_2}\cdots V S^{\ell_{m-1}}VP_0\label{eq:amdefinition}\\
\cU^{(m)}&=\sum_{(\ell_1,\ldots,\ell_m)\in \cP_m}S^{\ell_1}VS^{\ell_2}V\cdots VS^{\ell_m}VP_0\ ,\label{eq:Useries}
\end{align}
where the ``reduced resolvent'' is
\begin{align}
S^\ell &=\begin{cases}
\sum_{j\neq 0} (-E_j^{(0)})^{-\ell}P_j\qquad&\textrm{ if }\ell>0\\
-P_0 &\textrm{ if }\ell=0\ 
\end{cases}\label{eq:reducedresolvent}
\end{align}
and  $\cP_{m}$ is the set of $m$-tuples of nonnegative integers defined as
\begin{align}
\cP_{m}=\left\{(\ell_1,\ldots,\ell_m)\ \Big|\ \sum_{i=1}^m \ell_i=m,\  \sum_{i=1}^p\ell_i\geq p\qquad\textrm{ for all }p=1,\ldots,m-1
\right\}\ .\label{eq:cpmdef}
\end{align} 
The convergence of the series~\eqref{eq:powerseriesexpansions}
 can be analyzed by bounding the norm of $\cU$ using the triangle inequality (The convergence of the series for $\cA$ in~\eqref{eq:powerseriesexpansions} follows from that of the series for $\cU$ because of the identity $\cA=\lambda P_0 V\cU$.), i.e.,
\begin{align}
\|\cU\| & \leq \sum_{m=0} ^\infty \lambda^m \|\cU^{(m)}\|\ .\label{eq:Unormseries}
\end{align}
This sum can further be bounded by inserting~\eqref{eq:Useries}
\begin{align}
\|\cU^{(m)}\| &\leq |\cP_m|\cdot \max_{(\ell_1,\ldots,\ell_m)\in\cP_m} \|S^{\ell_1} VS^{\ell_2}V\cdots V S^{\ell_m}VP_0\|\nonumber\\ 
&\leq 4^m \max_{(\ell_1,\ldots,\ell_m)\in\cP_m} \|S^{\ell_1} VS^{\ell_2}V\cdots V S^{\ell_m}VP_0\|\ .\label{eq:sumuconvergence}
\end{align}
 Applying the submultiplicativity property $\|AB\|\leq \|A\|\cdot\|B\|$ and the bound $\|S\|\leq \left(E_1^{(0)}\right)^{-1}$ immediately gives
\begin{align}
\max_{(\ell_1,\ldots,\ell_m)\in\cP_m} \|S^{\ell_1} VS^{\ell_2}V\cdots V S^{\ell_m}VP_0\| &\leq \left(\frac{\|V\|}{E_1^{(0)}}\right)^m\ .\label{eq:maxnormSprod}
\end{align}
Reinserting~\eqref{eq:maxnormSprod} and~\eqref{eq:sumuconvergence} into~\eqref{eq:Unormseries} leads to the well-known sufficient condition~\cite{jordanfarhi}
\begin{align}
\lambda<\frac{E_1^{(0)}}{4\|V\|}\  \label{eq:simplecriterion}
\end{align}
for convergence. In our case, however, we will obtain a more refined condition by directly bounding the expression in~\eqref{eq:sumuconvergence}.

\subsection{The clock-gadget: Perturbative gadgets for ordered products\label{sec:perturbativegadgetsordered}}
In this section, we show how to obtain an ordered product $M_0\cdots M_{\maxt-1}$ of operators as the low-energy effective Hamiltonian of a $2$-local Hamiltonian. Our main result is the following statement.

\vspace*{12pt} 
\noindent 
\begin{theorem}\label{thm:maingadget}
Let $H_0$ be a Hamiltonian on a bipartite system $\cH_S\otimes\cH_\countsys$ of the form $H_0=-P_0$, where $P_0=\Gamma_0\otimes\proj{0}_\countsys$ is the projection onto the ground space. Here $\{\ket{\counterv}\}_{\counterv=0}^{\maxt-1}$ is an orthonormal basis of $\cH_\countsys\cong\mathbb{C}^\maxt$. We identify $\ket{\maxt}\equiv\ket{0}$. Let 
\begin{align*}
V&=\sum_{\counterv=0}^{\maxt-1} \left(M_\counterv^\dagger \otimes \ket{\counterv+1}\bra{\counterv}+M_\counterv\otimes\ket{\counterv}\bra{\counterv+1}\right)\ ,
\end{align*} 
where we assume that the operators $\{M_\counterv\}_{\counterv}$ satisfy the following proportionality constraints (we write $A\propto B$ if there exists a scalar~$c$ such that $A=cB$):
\begin{enumerate}[(i)]
\item\label{it:gadgetfirst}
$M_\counterv M_{\counterv+1}M_{\counterv+1}^\dagger M_\counterv^\dagger \propto (M_\counterv M_\counterv^\dagger)^2$
\item \label{it:gadgetsecond}
$(M_\counterv M_\counterv^\dagger)^2\propto M_\counterv M_\counterv^\dagger$
\item\label{it:gadgetthird}
$M_\counterv M_\counterv^\dagger\propto M_{\counterv-1}^\dagger M_{\counterv-1}$
\item\label{it:gadgetfourth}
$\Gamma_0 M_0M_0^\dagger \Gamma_0\propto \Gamma_0$
\end{enumerate}
for all~$i$. Consider the Hamiltonian $H=H_0+\lambda V$. Let $\tilde{H}_{\textrm{eff}}(\Delta)$ be the shifted effective Hamiltonian as explained in Section~\ref{sec:blochperturbation}. 
There exists a function $f(\lambda)$ such that 
for  
\begin{align}
\lambda  < \frac{1 }{16\max_i \|M_i\|}\ \label{eq:convergenceconditioncounter}
\end{align}
the effective Hamiltonian is
\begin{align*}
\tilde{H}_{\textrm{eff}}(f(\lambda)) &=(-1)^{\maxt-1}\lambda^\maxt  H_{\textrm{target}}\otimes \proj{0}_\countsys
+ O(\lambda^{\maxt+1})\ ,
\end{align*}
where the target Hamiltonian is defined by
\begin{align}
H_{\textrm{target}}&=\Gamma_0 M_0\cdots M_{\maxt-1}\Gamma_0+\Gamma_0 M_{\maxt-1}^\dagger \cdots M_0^\dagger \Gamma_0\ .\label{eq:targethamiltonian}
\end{align}
\end{theorem}
\vspace*{12pt}

\noindent We point out that~\eqref{it:gadgetfirst}--\eqref{it:gadgetfourth} are trivially satisfied if every $M_i$ is a unitary.

We stress that the convergence condition~\eqref{eq:convergenceconditioncounter} does not depend on~$n$.  This is in sharp contrast to the simple sufficient condition~\eqref{eq:simplecriterion} which generally requires the gap to scale with the system size. In particular, we can perturbatively obtain $n$-local interactions from $2$-local ones without such unfavorable scaling. However, this comes at the cost of using an $n$-dimensional clock system.

The remainder of this section is devoted to the derivation of this theorem. Since we can neglect constant energy shifts, we will consider the shifted Hamiltonian
\begin{align*}
H_0=\id_{S\countsys}-P_0
\end{align*}
instead of $-P_0$ (as in the theorem)
in order to be able to apply the formulas of Section~\ref{sec:blochperturbation}.  We first  reorganize the expression~\eqref{eq:amdefinition}. The projection onto the complement of the ground space (the $(E=1)$-eigenspace) is 
\begin{align}P_1=P_0^\bot=\id_{S\countsys}-\Gamma_0\otimes\proj{0}_\countsys=(\id_S-\Gamma_0)\otimes\proj{0}_\countsys+\id_S\otimes(\id_\countsys-\proj{0}_\countsys)\label{eq:p1projection}
\end{align}

 The reduced resolvent-operators $S^\ell$ are given by~(cf.~\eqref{eq:reducedresolvent})
\begin{align*}
S^\ell &=\begin{cases}
(-1)^\ell P_1\qquad&\textrm{ if }\ell>0\\
-P_0 &\textrm{ if }\ell=0\ .
\end{cases}
\end{align*}
Up to a sign, the operators $P_0 V S^{\ell_1}V S^{\ell_2}\cdots V S^{\ell_{m-1}}VP_0$
therefore only depend on the set of indices $i$ for which $\ell_i$ is non-zero. In particular, we can rewrite~\eqref{eq:amdefinition} as
\begin{align*}
\cA^{(m)}&=\sum_{\varepsilon=(\varepsilon_1,\ldots,\varepsilon_{m-1})\in\{0,1\}^{m-1} }g_m(\varepsilon)P_0 V P_{\varepsilon_1}VP_{\varepsilon_2}\cdots VP_{\varepsilon_{m-1}}VP_0\ ,
\end{align*}
where
\begin{align}
g_m(\varepsilon) &=(-1)^{m-1}\sum_{\substack{(\ell_1,\ldots,\ell_{m-1})\in \cP_{m-1}\\
\{i|\ell_i\neq 0\}=\{i|\varepsilon_i\neq 0\} 
}}(-1)^{\sum_i (\varepsilon_i+\ell_i)}\ .\label{eq:gfunctiondef}
\end{align}
Let us focus on a term 
of the form $P_0V P_{\varepsilon_1}VP_{\varepsilon_2}\cdots VP_{\varepsilon_{m-1}}VP_0$. We define the operators~$W_\counterv^\dagger=M_\counterv^\dagger\otimes\ket{\counterv+1}\bra{\counterv}_\countsys$. We can then write the perturbation as
\begin{align*}
V=\sum_\tau Z_\tau\ ,
\end{align*}
where we index the family of operators $\cup_i \{W_i,W_i^\dagger\}=\{Z_\tau\}_\tau$ by $\tau$.  The expression of interest takes the form
\begin{align*}
P_0V P_{\varepsilon_1}VP_{\varepsilon_2}\cdots VP_{\varepsilon_{m-1}}VP_0&=\sum_{\tau_1,\ldots,\tau_m}P_0 Z_{\tau_1}P_{\varepsilon_1}Z_{\tau_2}P_{\varepsilon_2}\cdots P_{\varepsilon_{m-1}}Z_{\tau_m} P_0\ .
\end{align*}
A similar computation can be performed for the operators $\cU^{(m)}$.
Let us summarize what we obtained so far:
\vspace*{12pt}
\noindent
\begin{lemma}\label{lem:Ausefulform}
The operator $\cA^{(m)}$ (cf.~\eqref{eq:amdefinition})  is a linear combination 
\begin{align}
\cA^{(m)}&=\sum_{\varepsilon,{\bf Y}} g_m(\varepsilon)\Theta(\varepsilon,{\bf Y})\ ,\label{eq:Asum}
\end{align}
where the sum is over all $\varepsilon=(\varepsilon_1,\ldots,\varepsilon_{m-1})\in\{0,1\}^{m-1}$, ${\bf Y}=(Y_1,\ldots,Y_m)\in\left(\cup_i \{W_i,W_i^\dagger\}\right)^m$, the function $g_m$ is given by~\eqref{eq:gfunctiondef} and where
\begin{align}
\Theta(\varepsilon,{\bf Y})&=P_0Y_1P_{\varepsilon_1}Y_2\cdots P_{\varepsilon_{m-1}}Y_m P_0\label{eq:Thetabydef}
\end{align}
Similarly, we have
\begin{align}
\cU^{(m)}&= \sum_{\varepsilon,{\bf Y}}g_{m+1}(\varepsilon) \Gamma(\varepsilon,{\bf Y})\ ,\label{eq:uexprexplicit}
\end{align}
with the sum over all $\varepsilon=(\varepsilon_1,\ldots,\varepsilon_m)\in\{0,1\}^m$, ${\bf Y}=(Y_1,\ldots,Y_m)\in \left(\cup_i \left\{W_i,W_i^\dagger\right\}\right)^m$ and 
\begin{align*}
\Gamma(\varepsilon,{\bf Y}) &= P_{\varepsilon_1} Y_1 P_{\varepsilon_2} Y_2\cdots Y_{m-1} P_{\varepsilon_m}Y_m P_0\ .
\end{align*}
\end{lemma}
\vspace*{12pt}
\noindent

Note that the operators $\Theta(\varepsilon,{\bf Y})$ constituting $\cA^{(m)}$ contain $m$~factors $Y_i$;  we will refer to this as operator of order~$m$. Our main technical result is a characterization of the operators $\Theta(\varepsilon,{\bf Y})$  of order $m\leq n$. Two operators of order~$n$~play a special role; these are
\begin{align}
\Theta_n^{\downarrow}&=\Theta(\underbrace{(1,\ldots,1)}_{n-1},(W_0,\ldots,W_{n-1}))\qquad\qquad\Theta_n^{\uparrow}=\Theta(\underbrace{(1,\ldots,1)}_{n-1},(W_{n-1}^\dagger,\ldots,W_0^\dagger))\ .\label{eq:updowndiagrams}
\end{align}
\vspace*{12pt}
\noindent
\begin{lemma}\label{lem:centralproportionality}
Let $\Theta_n^{\downarrow},\Theta_n^{\uparrow}$ be defined by~\eqref{eq:updowndiagrams}. 
\begin{enumerate}[(a)]
\item\label{eq:atpropto}
All operators $\Theta(\varepsilon,{\bf Y})\not\in\{\Theta_n^{\downarrow},\Theta_n^{\uparrow}\}$ of order $m\leq n$ satisfy $\Theta(\varepsilon,{\bf Y})\propto P_0$. 
\item\label{it:simplelemmasecondclaim}
$\Theta^{\downarrow}_n=\Gamma_0 M_0\cdots M_{n-1}\Gamma_0\otimes\proj{0}_I$ and 
$\Theta^{\uparrow}_n =\Gamma_0 M_{n-1}^\dagger\cdots M_0^\dagger\Gamma_0\otimes\proj{0}_I$.
\end{enumerate}
\end{lemma}
\vspace*{12pt}
\noindent
For the proof of Lemma~\ref{lem:centralproportionality}, observe that $W_\counterv^\dagger$ increases the counter variable $\countsys$ from~$\counterv$ to~$\counterv+1$, whereas $W_\counterv$ decreases it from~$\counterv+1$ to $\counterv$. In particular, all terms that do not match the zeros in the sequence $(0,\varepsilon_1,\ldots,\varepsilon_{m-1},0)$ vanish. For example, we have (for $\maxt=5$)
\begin{align*}
\Theta(\cdot,W_i)&=P_0W_iP_0=0\\
\Theta(\cdot,W_i^\dagger )&=P_0W_i^\dagger P_0=0\qquad\textrm{that is}\qquad\Theta(\cdot,{\bf Y})=0\qquad\textrm{ for all }{\bf Y}=Y_1\\
\Theta(11,{\bf Y})&=0\qquad\textrm{ for all }{\bf Y}=(Y_1,Y_2,Y_3)\\
\sum_{\bf Y} \Theta(1,{\bf Y}) &=P_0VP_1VP_0= P_0W_0P_1W_0^\dagger P_0+  P_0 W_{n-1}^\dagger P_1W_{n-1} P_0\\
\sum_{\bf Y} \Theta(1111,{\bf Y})&= P_0VP_1VP_1VP_1V P_1V P_0 = P_0 W_4^\dagger P_1W_3^\dagger P_1W_2^\dagger P_1 W_1^\dagger P_1W_0^\dagger P_0+h.c.\ . 
\end{align*}
We represent an operator $\Theta(\varepsilon,{\bf Y})$ of order~$m$ (cf.~\eqref{eq:Thetabydef}) by a diagram as follows. We first encode the sequence $0\varepsilon_1\ldots \varepsilon_{m-1}0$ by circles placed on the horizontal axis: $0$ is represented by a gray circle, and $1$~is represented by a black circle.
For example,\psset{unit=0.3cm}
\begin{align*}
\varepsilon_1\ldots \varepsilon_5=01100\rightarrow \raisebox{-0.25cm}{\begin{pspicture}(0,-1)(6,1)
\psgrid[subgriddiv=1,linecolor=lightgray,griddots=5,gridlabels=5pt](0,-1)(6,1)
\rput(0,0){\zzerobig}
\rput(1,0){\zzero}
\rput(2,0){\zone}
\rput(3,0){\zone}
\rput(4,0){\zzero}
\rput(5,0){\zzero}
\rput(6,0){\zzerobig}
\end{pspicture}}
\end{align*}
To encode the sequence ${\bf Y}$, we begin at the point $(m,0)$ on the horizontal axis and draw an upward-pointing arrow (vector $(-1,1)$) from the $\counterv$-th to the $(\counterv+1)$-th line for an operator $W_\counterv^\dagger$, and similarly a downward-pointing arrow starting from the $(\counterv+1)$-th line to the $\counterv$-th line for $W_\counterv$ (vector $(-1,-1)$). 
Interpreting the vertical coordinates modulo~$\maxt$, we have for example
\begin{align*}
\Theta(1,(W_0,W_0^\dagger))=P_0W_0 P_1 W_0^\dagger P_0 &=\qquad\raisebox{-0.25cm}{\begin{pspicture}(0,-1)(2,1)
\psgrid[subgriddiv=1,linecolor=lightgray,griddots=5,gridlabels=5pt](0,-1)(2,1)
\rput(0,0){\zzerobig}
\rput(1,0){\zone}
\rput(2,0){\zzerobig}
\rput(0,0){\zdown}
\rput(1,1){\zup}
\end{pspicture}}\\
\Theta(1,(W_{\maxt-1}^\dagger,W_{\maxt-1}))=P_0W_{\maxt-1}^\dagger P_1 W_{\maxt-1} P_0 &=\qquad\raisebox{-0.25cm}{\begin{pspicture}(0,-1)(2,1)
\psgrid[subgriddiv=1,linecolor=lightgray,griddots=5,gridlabels=5pt](0,-1)(2,1)
\rput(0,0){\zzerobig}
\rput(1,0){\zone}
\rput(2,0){\zzerobig}
\rput(0,0){\zup}
\rput(1,-1){\zdown}
\end{pspicture}}
\end{align*}
and  for $\maxt=5$
\begin{align}
\Theta_5^{\downarrow}&=\raisebox{-0.73cm}{\begin{pspicture}(0,0)(5,5)
\psgrid[subgriddiv=1,linecolor=lightgray,griddots=5,gridlabels=5pt](0,0)(5,5)
\rput(0,0){\zzerobig}\rput(1,0){\zone}\rput(2,0){\zone}\rput(3,0){\zone}\rput(4,0){\zone}\rput(5,0){\zzerobig}
\rput(0,5){\zzerobig}\rput(1,5){\zone}\rput(2,5){\zone}\rput(3,5){\zone}\rput(4,5){\zone}\rput(5,5){\zzerobig}
\rput(0,0){\zdown}
\rput(1,1){\zdown}
\rput(2,2){\zdown}
\rput(3,3){\zdown}
\rput(4,4){\zdown}
\end{pspicture}}\qquad\qquad
\Theta_5^{\uparrow}=
\raisebox{-0.73cm}{\begin{pspicture}(0,0)(5,5)
\psgrid[subgriddiv=1,linecolor=lightgray,griddots=5,gridlabels=5pt](0,0)(5,5)
\rput(0,0){\zzerobig}\rput(1,0){\zone}\rput(2,0){\zone}\rput(3,0){\zone}\rput(4,0){\zone}\rput(5,0){\zzerobig}
\rput(0,5){\zzerobig}\rput(1,5){\zone}\rput(2,5){\zone}\rput(3,5){\zone}\rput(4,5){\zone}\rput(5,5){\zzerobig}
\rput(0,5){\zup}
\rput(1,4){\zup}
\rput(2,3){\zup}
\rput(3,2){\zup}
\rput(4,1){\zup}
\end{pspicture}}\label{eq:diagonaldiagramspicture}
\end{align}
By definition of the operators $\Theta(\varepsilon,{\bf Y})$, only diagrams  \begin{enumerate}
\item whose sequence of arrows forms a continuous path which
\item starts at $(m,0)$ (modulo $(0,n)$),
\item ends at $(0,0)$ (modulo $(0,n)$) and
\item goes through the points $(\varepsilon_\counterv,0)$ (modulo $(0,n)$) for all~$\counterv$ with $\varepsilon_i=0$ 
\end{enumerate}
correspond to a non-zero operator $\Theta(\varepsilon,{\bf Y})$ of order $m$.  We call such diagrams  valid.  In other words, in every valid diagram the sequence of arrows must form a continuous path  which passes through every gray circle. For example, $\cA^{(4)}$ is the sum of the following valid diagrams:
\begin{align*}
\cA^{(4)}&=\qquad
g(111)\cdot \left(\raisebox{-0.55cm}{\begin{pspicture}(0,-2)(4,2)
\psgrid[subgriddiv=1,linecolor=lightgray,griddots=5,gridlabels=5pt](0,-2)(4,2)
\rput(0,0){\zzerobig}\rput(1,0){\zone}\rput(2,0){\zone}\rput(3,0){\zone}\rput(4,0){\zzerobig}
\rput(0,0){\zdown}
\rput(1,1){\zdown}
\rput(2,2){\zup}
\rput(3,1){\zup}
\end{pspicture}}
\ +\ 
\raisebox{-0.55cm}{\begin{pspicture}(0,-2)(4,2)
\psgrid[subgriddiv=1,linecolor=lightgray,griddots=5,gridlabels=5pt](0,-2)(4,2)
\rput(0,0){\zzerobig}\rput(1,0){\zone}\rput(2,0){\zone}\rput(3,0){\zone}\rput(4,0){\zzerobig}
\rput(0,0){\zup}
\rput(1,-1){\zup}
\rput(2,-2){\zdown}
\rput(3,-1){\zdown}
\end{pspicture}}
\ +\ 
\raisebox{-0.55cm}{\begin{pspicture}(0,-2)(4,2)
\psgrid[subgriddiv=1,linecolor=lightgray,griddots=5,gridlabels=5pt](0,-2)(4,2)
\rput(0,0){\zzerobig}\rput(1,0){\zone}\rput(2,0){\zone}\rput(3,0){\zone}\rput(4,0){\zzerobig}
\rput(0,0){\zdown}
\rput(1,1){\zup}
\rput(2,0){\zdown}
\rput(3,1){\zup}
\end{pspicture}}
\ +\ 
\raisebox{-0.55cm}{\begin{pspicture}(0,-2)(4,2)
\psgrid[subgriddiv=1,linecolor=lightgray,griddots=5,gridlabels=5pt](0,-2)(4,2)
\rput(0,0){\zzerobig}\rput(1,0){\zone}\rput(2,0){\zone}\rput(3,0){\zone}\rput(4,0){\zzerobig}
\rput(0,0){\zup}
\rput(1,-1){\zdown}
\rput(2,0){\zup}
\rput(3,-1){\zdown}
\end{pspicture}}
\ +\ 
\raisebox{-0.55cm}{\begin{pspicture}(0,-2)(4,2)
\psgrid[subgriddiv=1,linecolor=lightgray,griddots=5,gridlabels=5pt](0,-2)(4,2)
\rput(0,0){\zzerobig}\rput(1,0){\zone}\rput(2,0){\zone}\rput(3,0){\zone}\rput(4,0){\zzerobig}
\rput(0,0){\zdown}
\rput(1,1){\zup}
\rput(2,0){\zup}
\rput(3,-1){\zdown}
\end{pspicture}}
\ +\ 
\raisebox{-0.55cm}{\begin{pspicture}(0,-2)(4,2)
\psgrid[subgriddiv=1,linecolor=lightgray,griddots=5,gridlabels=5pt](0,-2)(4,2)
\rput(0,0){\zzerobig}\rput(1,0){\zone}\rput(2,0){\zone}\rput(3,0){\zone}\rput(4,0){\zzerobig}
\rput(0,0){\zup}
\rput(1,-1){\zdown}
\rput(2,0){\zdown}
\rput(3,1){\zup}
\end{pspicture}}
\right)\\
&\qquad +g(101)\left(\raisebox{-0.55cm}{\begin{pspicture}(0,-2)(4,2)
\psgrid[subgriddiv=1,linecolor=lightgray,griddots=5,gridlabels=5pt](0,-2)(4,2)
\rput(0,0){\zzerobig}\rput(1,0){\zone}\rput(2,0){\zzero}\rput(3,0){\zone}\rput(4,0){\zzerobig}
\rput(0,0){\zdown}
\rput(1,1){\zup}
\rput(2,0){\zdown}
\rput(3,1){\zup}
\end{pspicture}}
\qquad+\qquad
\raisebox{-0.55cm}{\begin{pspicture}(0,-2)(4,2)
\psgrid[subgriddiv=1,linecolor=lightgray,griddots=5,gridlabels=5pt](0,-2)(4,2)
\rput(0,0){\zzerobig}\rput(1,0){\zone}\rput(2,0){\zzero}\rput(3,0){\zone}\rput(4,0){\zzerobig}
\rput(0,0){\zup}
\rput(1,-1){\zdown}
\rput(2,0){\zup}
\rput(3,-1){\zdown}
\end{pspicture}}
\qquad + \qquad
\raisebox{-0.55cm}{\begin{pspicture}(0,-2)(4,2)
\psgrid[subgriddiv=1,linecolor=lightgray,griddots=5,gridlabels=5pt](0,-2)(4,2)
\rput(0,0){\zzerobig}\rput(1,0){\zone}\rput(2,0){\zzero}\rput(3,0){\zone}\rput(4,0){\zzerobig}
\rput(0,0){\zdown}
\rput(1,1){\zup}
\rput(2,0){\zup}
\rput(3,-1){\zdown}
\end{pspicture}}
\qquad+\qquad
\raisebox{-0.55cm}{\begin{pspicture}(0,-2)(4,2)
\psgrid[subgriddiv=1,linecolor=lightgray,griddots=5,gridlabels=5pt](0,-2)(4,2)
\rput(0,0){\zzerobig}\rput(1,0){\zone}\rput(2,0){\zzero}\rput(3,0){\zone}\rput(4,0){\zzerobig}
\rput(0,0){\zup}
\rput(1,-1){\zdown}
\rput(2,0){\zdown}
\rput(3,1){\zup}
\end{pspicture}}\right)
\end{align*}
Using~\eqref{eq:p1projection}, we can simplify the expressions for operators corresponding to such diagrams. That is, we use the fact that $P_1$ acts trivially on the subspace $\cH_S\otimes\mathsf{span}\{\ket{\counterv}_\countsys\}_{\counterv\neq 0}$, and acts as $(\id-\Gamma_0)\otimes\proj{0}_\countsys$ otherwise. This gives for example
\begin{align*}
P_0W_0P_1W_0^\dagger P_0&=\Gamma_0 M_0M_0^\dagger \Gamma_0\otimes\proj{0}_\countsys\\
P_0 W_4^\dagger P_1W_3^\dagger P_1W_2^\dagger P_1 W_1^\dagger P_1W_0^\dagger P_0&=\Gamma_0 M_4^\dagger M_3^\dagger M_2^\dagger M_1^\dagger M_0^\dagger \Gamma_0\otimes\proj{0}_\countsys\ .
\end{align*}
Essentially substituting $W_\counterv$ by $M_\counterv$, $P_1$ by the identity  (or $\id-\Gamma_0$) and $P_0$ by $\Gamma_0$, 
we can use the same pictorial representation for the resulting product of  operators $\{M_\counterv\}_\counterv$. Special care needs to be taken in cases where the arrows touch the horizontal axis: If the corresponding circle is black, the operator $\Gamma_1=\id-\Gamma_0$ needs to be inserted because of~\eqref{eq:p1projection}; if it is gray, we need to insert $\Gamma_0$. Again, examples are 
\begin{align*}
\Gamma_0 M_0 M_0^\dagger \Gamma_0 &=\ \raisebox{-0.25cm}{\begin{pspicture}(0,-1)(2,1)
\psgrid[subgriddiv=1,linecolor=lightgray,griddots=5,gridlabels=5pt](0,-1)(2,1)
\rput(0,0){\zzerobig}
\rput(1,0){\zone}
\rput(2,0){\zzerobig}
\rput(0,0){\zdown}
\rput(1,1){\zup}
\end{pspicture}}\\
\Gamma_0 M_{n-1}^\dagger M_{n-1}\Gamma_0M_0M_0^\dagger\Gamma_0 &=\ \raisebox{-0.55cm}{\begin{pspicture}(0,-2)(4,2)
\psgrid[subgriddiv=1,linecolor=lightgray,griddots=5,gridlabels=5pt](0,-2)(4,2)
\rput(0,0){\zzerobig}\rput(1,0){\zone}\rput(2,0){\zzero}\rput(3,0){\zone}\rput(4,0){\zzerobig}
\rput(0,0){\zup}
\rput(1,-1){\zdown}
\rput(2,0){\zdown}
\rput(3,1){\zup}
\end{pspicture}}\\
\Gamma_0 M_{n-1}^\dagger M_{n-1}\Gamma_1M_0M_0^\dagger\Gamma_0 &=\ \raisebox{-0.55cm}{\begin{pspicture}(0,-2)(4,2)
\psgrid[subgriddiv=1,linecolor=lightgray,griddots=5,gridlabels=5pt](0,-2)(4,2)
\rput(0,0){\zzerobig}\rput(1,0){\zone}\rput(2,0){\zone}\rput(3,0){\zone}\rput(4,0){\zzerobig}
\rput(0,0){\zup}
\rput(1,-1){\zdown}
\rput(2,0){\zdown}
\rput(3,1){\zup}
\end{pspicture}}
\end{align*} 
In particular, the assumptions of Theorem~\ref{thm:maingadget} take the form
\begin{enumerate}[(i)]
\item
$\raisebox{-0.25cm}{\begin{pspicture}(0,0)(4,2)
\psgrid[subgriddiv=1,linecolor=lightgray,griddots=5,gridlabels=0pt](0,0)(4,2)
\rput(0,0){\zdown}
\rput(1,1){\zdown}
\rput(2,2){\zup}
\rput(3,1){\zup}
\end{pspicture}}\propto
\raisebox{-0.25cm}{\begin{pspicture}(0,0)(4,2)
\psgrid[subgriddiv=1,linecolor=lightgray,griddots=5,gridlabels=0pt](0,0)(4,2)
\rput(0,0){\zdown}
\rput(1,1){\zup}
\rput(2,0){\zdown}
\rput(3,1){\zup}
\end{pspicture}}$
\item
$\raisebox{-0.125cm}{\begin{pspicture}(0,0)(4,1)
\psgrid[subgriddiv=1,linecolor=lightgray,griddots=5,gridlabels=0pt](0,0)(4,1)
\rput(0,0){\zdown}
\rput(1,1){\zup}
\rput(2,0){\zdown}
\rput(3,1){\zup}
\end{pspicture}}\propto
\raisebox{-0.125cm}{\begin{pspicture}(0,0)(2,1)
\psgrid[subgriddiv=1,linecolor=lightgray,griddots=5,gridlabels=0pt](0,0)(2,1)
\rput(0,0){\zdown}
\rput(1,1){\zup}
\end{pspicture}}$
\item
$\raisebox{-0.25cm}{\begin{pspicture}(0,0)(2,2)
\psgrid[subgriddiv=1,linecolor=lightgray,griddots=5,gridlabels=0pt](0,0)(2,2)
\rput(0,1){\zdown}
\rput(1,2){\zup}
\end{pspicture}}\propto
\raisebox{-0.25cm}{\begin{pspicture}(0,0)(2,2)
\psgrid[subgriddiv=1,linecolor=lightgray,griddots=5,gridlabels=0pt](0,0)(2,2)
\rput(0,1){\zup}
\rput(1,0){\zdown}
\end{pspicture}}$
\item
$\raisebox{-0.25cm}{\begin{pspicture}(0,-1)(2,1)
\psgrid[subgriddiv=1,linecolor=lightgray,griddots=5,gridlabels=0pt](0,-1)(2,1)
\rput(0,0){\zzero}\rput(1,0){\zone}\rput(2,0){\zzero}
\rput(0,0){\zdown}
\rput(1,1){\zup}
\end{pspicture}}\propto 
\raisebox{-0.10cm}{
\begin{pspicture}(-0.5,-0.5)(0.5,0.5)
\psgrid[subgriddiv=1,linecolor=lightgray,griddots=5,gridlabels=0pt](0,-1)(0.5,1)\rput(0,0){\zzero}
\end{pspicture}}\propto
\Gamma_0$\ , 
\end{enumerate}
where the absence of circles indicates that we are away from the horizontal axis. Let us verify that these rules also hold for the horizontal axis. The analog of~\eqref{it:gadgetfirst}, combined with~\eqref{it:gadgetsecond} is
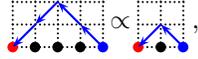
\begin{align*}
\raisebox{-0.25cm}{\begin{pspicture}(0,0)(4,2)
\psgrid[subgriddiv=1,linecolor=lightgray,griddots=5,gridlabels=0pt](0,0)(4,2)
\rput(0,0){\zundefined}\rput(1,0){\zone}\rput(2,0){\zone}\rput(3,0){\zone}\rput(4,0){\zundefinedsecond}
\rput(0,0){\zdown}
\rput(1,1){\zdown}
\rput(2,2){\zup}
\rput(3,1){\zup}
\end{pspicture}}\propto
\raisebox{-0.25cm}{\begin{pspicture}(0,0)(2,2)
\psgrid[subgriddiv=1,linecolor=lightgray,griddots=5,gridlabels=0pt](0,0)(2,2)
\rput(0,0){\zundefined}\rput(1,0){\zone}\rput(2,0){\zundefinedsecond}
\rput(0,0){\zdown}
\rput(1,1){\zup}
\end{pspicture}}\ ,
\end{align*}
where the (red and blue) circles at the endpoints are arbitrary. This directly follows from assumption~\eqref{it:gadgetfirst} and \eqref{it:gadgetsecond}, i.e., the fact that $M_0M_1M_1^\dagger M_0^\dagger \propto (M_0M_0^\dagger)^2\propto M_0 M_0^\dagger$.

Next we show an analog of~\eqref{it:gadgetsecond}, which is
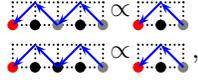
\begin{align*}
\raisebox{-0.125cm}{\begin{pspicture}(0,0)(4,1)
\psgrid[subgriddiv=1,linecolor=lightgray,griddots=5,gridlabels=0pt](0,0)(4,1)
\rput(0,0){\zundefined}\rput(1,0){\zone}\rput(2,0){\zzero}\rput(3,0){\zone}\rput(4,0){\zzero}
\rput(0,0){\zdown}
\rput(1,1){\zup}
\rput(2,0){\zdown}
\rput(3,1){\zup}
\end{pspicture}}&\propto
\raisebox{-0.125cm}{\begin{pspicture}(0,0)(2,1)
\psgrid[subgriddiv=1,linecolor=lightgray,griddots=5,gridlabels=0pt](0,0)(2,1)
\rput(0,0){\zundefined}\rput(1,0){\zone}\rput(2,0){\zzero}
\rput(0,0){\zdown}
\rput(1,1){\zup}
\end{pspicture}}\\
\raisebox{-0.125cm}{\begin{pspicture}(0,0)(4,1)
\psgrid[subgriddiv=1,linecolor=lightgray,griddots=5,gridlabels=0pt](0,0)(4,1)
\rput(0,0){\zundefined}\rput(1,0){\zone}\rput(2,0){\zone}\rput(3,0){\zone}\rput(4,0){\zzero}
\rput(0,0){\zdown}
\rput(1,1){\zup}
\rput(2,0){\zdown}
\rput(3,1){\zup}
\end{pspicture}}&\propto
\raisebox{-0.125cm}{\begin{pspicture}(0,0)(2,1)
\psgrid[subgriddiv=1,linecolor=lightgray,griddots=5,gridlabels=0pt](0,0)(2,1)
\rput(0,0){\zundefined}\rput(1,0){\zone}\rput(2,0){\zzero}
\rput(0,0){\zdown}
\rput(1,1){\zup}
\end{pspicture}}\ ,
\end{align*}
where the (red) circle at the left endpoint can be arbitrary. The first identity is immediate from assumption~\eqref{it:gadgetsecond}. The second identity follows from
\begin{align*}
M_0M_0^\dagger\Gamma_1 M_0M_0^\dagger\Gamma_0&=(M_0M_0^\dagger)^2\Gamma_0-M_0M_0^\dagger (\Gamma_0 M_0M_0^\dagger \Gamma_0)\\
&=c_1\cdot M_0M_0^\dagger \Gamma_0+c_2 \cdot M_0M_0^\dagger \Gamma_0 
\end{align*}
where we used~\eqref{it:gadgetsecond} and~\eqref{it:gadgetfourth}. Finally,~\eqref{it:gadgetthird} takes the form
\begin{align*}
\raisebox{-0.25cm}{\begin{pspicture}(0,0)(2,2)
\psgrid[subgriddiv=1,linecolor=lightgray,griddots=5,gridlabels=0pt](0,0)(2,2)
\rput(0,1){\zundefined}\rput(1,1){\zone}\rput(2,1){\zundefinedsecond}
\rput(0,1){\zdown}
\rput(1,2){\zup}
\end{pspicture}}\propto
\raisebox{-0.25cm}{\begin{pspicture}(0,0)(2,2)
\psgrid[subgriddiv=1,linecolor=lightgray,griddots=5,gridlabels=0pt](0,0)(2,2)
\rput(0,1){\zundefined}\rput(1,1){\zone}\rput(2,1){\zundefinedsecond}
\rput(0,1){\zup}
\rput(1,0){\zdown}
\end{pspicture}}
\end{align*} 
with arbitrary endpoints. This is again an immediate consequence of assumption~\eqref{it:gadgetthird}. With these rules, we are ready to prove Lemma~\ref{lem:centralproportionality}.

\noindent{\bf Proof of Lemma~\ref{lem:centralproportionality}:}
It is easy to check that by repeated application of rules~\eqref{it:gadgetfirst}--\eqref{it:gadgetthird} and their extensions, every valid diagram corresponding to an operator~$\Theta(\varepsilon,{\bf Y})\not\in\{\Theta^\uparrow_n,\Theta^\downarrow_n\}$ of order $m\leq n$ can be reduced to the diagram in~\eqref{it:gadgetfourth}.  This implies the first claim~\eqref{eq:atpropto}. The second claim~\eqref{it:simplelemmasecondclaim} immediately follows from the definitions. \square\,

Theorem~\ref{thm:maingadget} is an immediate consequence of  Lemma~\ref{lem:centralproportionality}.

\noindent{\bf Proof of Theorem~\ref{thm:maingadget}:}
Combining~\eqref{eq:Asum} with Lemma~\ref{lem:centralproportionality} gives
\begin{align*}
\cA^{(m)} &\propto P_0\qquad\textrm{ for all }m<n\\
\cA^{(n)} &=const\cdot P_0+g_n(\underbrace{(1,\ldots,1)}_{n-1})\cdot \left(\Theta^\downarrow_n+\Theta^\uparrow_n\right)
\end{align*}
We conclude that \begin{align*}
\sum_{m=0}^{\maxt} \lambda^m \cA^{(m)}&=f(\lambda)\cdot P_0+g_n(\underbrace{(1,\ldots,1)}_{n-1})\lambda^\maxt \left(\Gamma_0 M_0\cdots M_{\maxt-1}\Gamma_0+\Gamma_0 M_{\maxt-1}^\dagger \cdots M_0^\dagger \Gamma_0\right)\otimes\proj{0}_\countsys\ .
\end{align*}
By~\eqref{eq:pzeroconstantshift}, the first summand only shifts the energy of the effective Hamiltonian $H_{\textrm{eff}}=\cU\cA\cU^\dagger$. Moreover, 
since the above identity is already a $\maxt$-th order approximation to $\cA$, it suffices
to use the $0$-th order approximation $\cU\approx \cU^{(0)}=P_0$ to compute $H_{\textrm{eff}}$ (cf.~\eqref{eq:auzeroorder}) to order $\maxt$. 
The expression for the effective Hamiltonian in the theorem follows because $g_n(\underbrace{(1,\ldots,1)}_{n-1})=(-1)^{n-1}$  according to~\eqref{eq:gfunctiondef}. 

To analyze the convergence of the perturbation series,  we use~\eqref{eq:uexprexplicit} derived in Lemma~\ref{lem:Ausefulform} and the triangle inequality to get
\begin{align*}
\|\cU^{(m)}\|&\leq \sum_\varepsilon |g_{m+1}(\varepsilon)| \cdot \sum_{{\bf Y}} \|\Gamma(\varepsilon,{\bf Y})\|\\
&\leq  2^m\cdot 2^m\max_\varepsilon |g_{m+1}(\varepsilon)|\cdot \max_{\bf Y} \|\Gamma(\varepsilon,{\bf Y})\|\ .
\end{align*}
Here we have used the fact that the sum over ${\bf Y}$ can be restricted to all valid paths, since the operator $\Gamma(\varepsilon,{\bf Y})$ is zero otherwise. Using the submultiplicativity property of the operator norm, we have 
\begin{align*}
\|\Gamma(\varepsilon,{\bf Y})\|&\leq \|Y_1\|\cdots \|Y_m\|\leq \left(\max_i\max \{ \|W_i\|, \|W_i^\dagger\|\}\right)^m=:\gamma^m\ .
\end{align*}
With the bound $|g_{m+1}(\varepsilon)|\leq |\cP_m|\leq 4^m$, we conclude that 
\begin{align*}
\|\cU^{(m)}\|\leq (16\gamma)^m\ .
\end{align*}
From~\eqref{eq:Unormseries} and the fact that the bounded operators form a Banach-$*$-algebra, that is,
\begin{align*}
\|W_i\|^2=\|W_i^\dagger\|^2 &= \|W_i W_i^\dagger\|=\Big\|M_i M_i^\dagger\otimes\proj{i}\Big\|=\|M_iM_i^\dagger\|=\|M_i\|^2\ ,
\end{align*}
we conclude that the bound given in the theorem is sufficient to guarantee convergence of the perturbation series. \square\,

\subsection{Extension to several clocks\label{sec:multiclock}}
In this section, we extend Theorem~\ref{thm:maingadget} to 
a situation where we have $L$~sets of operators $\left(\Gamma_0^\alpha,\{M^\alpha_i\}_{i=0}^{n-1}\right)$ indexed by $\alpha=1,\ldots,L$. To every index $\alpha$, we associate a target Hamiltonian $H^\alpha_{\textrm{target}}$ given by the analog of~\eqref{eq:targethamiltonian}, that is,
\begin{align}\label{eq:targethamiltonianseveral}
H^\alpha_{\textrm{target}}&=\Gamma M_0^\alpha\cdots M_{\maxt-1}^\alpha\Gamma+\Gamma (M_{\maxt-1}^\alpha)^\dagger \cdots (M_0^\alpha)^\dagger \Gamma\qquad\textrm{ where }\qquad \Gamma=\prod_{\alpha} \Gamma^\alpha_0\ .
\end{align}
We will assume that the operators $\{\Gamma^{\alpha}_0\}_\alpha$ are commuting projections, that is 
\begin{align}
(\Gamma^{\alpha}_0)^2=(\Gamma^{\alpha}_0)^\dagger=\Gamma^{\alpha}_0\qquad \textrm{and }\qquad [\Gamma^{\alpha}_0,\Gamma^{\beta}_0]=0\qquad\textrm{ for all }\alpha,\beta\ .\label{eq:gammamultdef}
\end{align}

Our aim is to construct a Hamiltonian whose low-energy effective Hamiltonian is equal to the sum of these target Hamiltonians. We will do so in a manner similar to Theorem~\ref{thm:maingadget}. In particular, we will assume that
for every fixed index~$\alpha$, the operators $\left(\Gamma_0^\alpha,\{M^\alpha_i\}_{i=0}^{n-1}\right)$ satisfy the conditions of Theorem~\ref{thm:maingadget}. That is, they obey the rules
\begin{enumerate}[(i)]
\item
$M_\counterv^\alpha M_{\counterv+1}^\alpha(M_{\counterv+1}^\alpha)^\dagger (M_\counterv^\alpha)^\dagger \propto (M_\counterv^\alpha (M_\counterv^\alpha)^\dagger)^2$\label{it:multicommfirst}
\item 
$(M_\counterv^\alpha (M_\counterv^\alpha)^\dagger)^2\propto M_\counterv^\alpha (M_\counterv^\alpha)^\dagger$\label{it:multicommsecond}
\item
$M_\counterv^\alpha(M_\counterv^\alpha)^\dagger\propto (M_{\counterv-1}^\alpha)^\dagger M_{\counterv-1}^\alpha$\label{it:multicommthird}
\item
$\Gamma_0^\alpha M_0^\alpha(M_0^\alpha)^\dagger \Gamma_0^\alpha\propto \Gamma_0^\alpha$\label{it:multicommfourth}
\end{enumerate}
for all $\alpha$ and $i$. It turns out that conditions~\eqref{it:multicommfirst}--\eqref{it:multicommfourth} are insufficient for our purposes; a complication arises because  operators 
with different indices $\alpha\neq\beta$ may not commute. In addition to the above conditions, we will therefore require the following commutation relations for all $\alpha\neq \beta$ and $i$:
\begin{align}
[M_i^\alpha,\Gamma_0^\beta]&=[(M_i^\alpha)^\dagger,\Gamma_0^\beta]=0\label{eq:secondcommutat}\\
[M_i^\alpha,(M_0^\beta)^\dagger M_0^\beta]&=[(M_i^\alpha)^\dagger,(M_0^\beta)^\dagger M_0^\beta]=0 \label{eq:specialcommutativity}
\end{align}
We stress that we do not require commutativity of the operators $M_i^\alpha$ in the form $[M_i^\alpha,M_j^\beta]=0$; indeed, these operators will in general not commute in applications of interest. We will take care of this non-commutativity by inserting additional operators into the perturbation. It will be convenient to define the sets of indices
\begin{align*}
\chi(\alpha,i) &=\{\beta\ | \beta\neq\alpha\textrm{ and } \exists j: [M^\alpha_i,M^\beta_j]\neq 0\textrm{ or } [M^\alpha_i,(M^\beta_j)^\dagger]\neq 0\}\  .
\end{align*}
To define our unperturbed Hamiltonian $H_0$ and the perturbation $V$, we will introduce $L$~auxiliary systems $\cH_{\countsys^1}\otimes\cdots\otimes \cH_{\countsys^L}$.
As before, we assume that $\cH_{I^\alpha}\cong \mathbb{C}^n$ with orthonormal basis $\{\ket{i}_\alpha\}_{i=0}^{n-1}$. To keep the expressions short, we will omit identities when clear from the context.

\vspace*{12pt}
\noindent
\newtheorem*{varthm}{Theorem~\ref{thm:maingadget}$^\prime$}
\begin{varthm}
Let $\left(\Gamma_0^\alpha,\{M^\alpha_i\}_{i=0}^{n-1}\right)$ with $\alpha=1,\ldots,L$ be a family of operators on a Hilbert space $\cH$ with  properties~\eqref{eq:gammamultdef}--\eqref{eq:specialcommutativity} and~\eqref{it:multicommfirst}--\eqref{it:multicommfourth}.  Let
\begin{align*}
H_0 &=-\sum_{\alpha}\Gamma_0^\alpha
\otimes\proj{0}_{\alpha}\ .
\end{align*}
be a Hamiltonian on $\cH\otimes \cH_{I_1}\otimes\cdots\otimes\cH_{I_L}$, where $\cH_{I_\alpha}\cong\mathbb{C}^n$ has orthonormal basis $\{\ket{i}_\alpha\}_{i=0}^{n-1}$. Let
\begin{align*}
V &=\sum_{\alpha,i} \left((M_i^\alpha)^\dagger\otimes\ket{i+1}\bra{i}_{\alpha}\otimes \cK(\alpha,i) + M_i^\alpha\otimes\ket{i}\bra{i+1}_{\alpha}\otimes \cK(\alpha,i)\right)\ ,
\end{align*}
where $\cK(\alpha,i)=\bigotimes_{\beta\in\chi(\alpha,i)}\proj{0}_\beta$ projects onto states whose register $\cH_{\cI^{\beta}}$ is in state $\ket{0}_\beta$ for all $\beta\in\chi(\alpha,i)$.  Consider the Hamiltonian $H=H_0+\lambda V$. Then there exists a function $f(\lambda)$ such that for sufficiently small~$\lambda$
\begin{align*}
\tilde{H}_{\textrm{eff}}(f(\lambda)) &=(-1)^{n-1} \lambda^n \left(\sum_\alpha H_{\textrm{target}}^\alpha\right)\otimes\proj{0}^{\otimes L}+O(\lambda^{n+1})\ ,
\end{align*}
where the target Hamiltonians $H_{\textrm{target}}^\alpha$ are given by~\eqref{eq:targethamiltonianseveral}.
\end{varthm}
\vspace*{12pt}
\noindent
We give a proof of this extension in appendix~\ref{sec:maingadgetvarproof}. It closely follows the proof of Theorem~\ref{thm:maingadget}, with an extended diagrammatic notation for several counters.

Note that we have not given a criterion for the convergence of the perturbation series in Theorem~\ref{thm:maingadget}. According to the naive condition~\eqref{eq:simplecriterion}, a scaling of the coupling strength as $\lambda \sim \frac{1}{L}$ (for constant~$n$) is sufficient. However, this condition is not entirely satisfactory, as~$L$ represents the system size.  A priori, it is unclear whether a significantly better criterion (possibly involving properties of the operators $\{M_i^\alpha\}_i$) can be found. This is an important open problem.

The extension from a single counter system (Theorem~\ref{thm:maingadget}) to several counter systems in Theorem~\ref{thm:maingadget}$^\prime$ is based on coupling non-commuting terms to certain projectors~(i.e., the terms $\hat{\chi}(\alpha,i)$) which ensure that the corresponding clocks are not simultaneously ``active'' (i.e., not in the state $\ket{0}$). This  allows us to treat the individual counters independently. The same behavior could potentially also be achieved by introducing terms in the Hamiltonian which  assign a high energy penalty to configurations that have more than one active clock. In fact, this may lead to more local terms. However, with this approach, the additional terms must 
have strong coupling  because the suppression of undesired configurations is only based on the energy denominators in the reduced resolvent. Analyzing whether this alternative approach yields useful results falls into the same category of problems as the previously mentioned one. A major difference is that 
 Theorem~\ref{thm:maingadget}$^\prime$ only involves one energy scale corresponding to the parameter~$\lambda$, whereas the method sketched here presumably requires at least two different energy scales in the Hamiltonian.

\section{Perturbative gadgets for quantum double models\label{sec:perturbativegadgetsquantumdouble}}
We are ready to apply the clock-gadget derived in Section~\ref{sec:perturbation} to the quantum double models discussed in Section~\ref{sec:nonabelian}. We first show how a single plaquette- or vertex-operator can be obtained perturbatively (Section~\ref{sec:singleoperator}); this is based on Proposition~\ref{prop:specialrep} and Theorem~\ref{thm:maingadget}. In Section~\ref{sec:allplaquettes}, we then use the more general Theorem~\ref{thm:maingadget}$^\prime$ to generate the full Hamiltonian.

\subsection{Generating a plaquette/vertex-operator\label{sec:singleoperator}}
 For concreteness, we consider the case of the honeycomb lattice~$\cL$ (it turns out that this case is slightly more involved than the case of a square lattice because the degrees of the vertices in the primal and the dual lattice are different.) We denote the Hilbert space of the qudits on the lattice by $\cH_{\cL}$.
\subsubsection{Generating a plaquette-term}
Consider a plaquette $p$. We give a construction of a Hamiltonian~$H_0^p$ and a perturbation~$V^p$ such that the effective Hamiltonian is proportional to the plaquette-operator~$B(p)$ on $\cH_\cL$.

For this purpose, we introduce two auxiliary systems associated to the plaquette $p$: a counter system $\cH_{I^p}\cong\mathbb{C}^6$ with orthonormal basis $\{\ket{i}\}_{i=0}^5$ and a system $\cH_{R^p}\cong \mathbb{C}^{|G|}$ with orthonormal basis $\{\ket{g}\}_{g\in G}$. The unperturbed Hamiltonian on $\cH_\cL\otimes\cH_{R^p}\otimes\cH_{I^p}$ is defined as
\begin{align}
H_0^p &=-\proj{\Psi^p}_{R^p}\otimes\proj{0}_{\countsys^p}\ ,\label{eq:hpzerodef}
\end{align}
where $\ket{\Psi^p}_{R^p}=\ket{1}$. To define the perturbation, let $e_0$, $e_1$,\ldots,$e_5$ be the edges on the boundary of $p$ in clockwise order and let
\begin{align*}
M_i^p=M(e_i,p) =\sum_{g} T^{g}(e_i,p)_{\cL}\otimes (L^g_+)_{R^p}\qquad\textrm{ for } i=0,\ldots,5
\end{align*}
 be the operators on $\cH_\cL\otimes \cH_{R^p}$ introduced in Proposition~\ref{prop:specialrep}. The perturbation  is
\begin{align}
V^p &=\sum_{i=0}^5\left((M_i^p)^\dagger\otimes \ket{i+1}\bra{i}_{I^p}+h.c.\right)\label{eq:vpdef}
\end{align}
Now consider the Hamiltonian $H^p=H_0^p+\lambda V^p$. We claim that the effective Hamiltonian is (up to a global energy shift~$f(\lambda)$) equal to
\begin{align}
\tilde{H}_{\textrm{eff}}^p(f(\lambda)) &=-2\lambda^6\cdot B(p)_\cL\otimes\proj{\Psi^p}_{R^p}\otimes\proj{0}_{I^p}+O(\lambda^{7})\ .\label{eq:effectivehamiltonianexpr}
\end{align}
To verify this statement, observe that the operators $\{M_i^p\}_i$ are unitary (cf.~Proposition~\ref{prop:specialrep}) and thus satisfy the conditions of Theorem~\ref{thm:maingadget}. Clearly, the projection  $P_0=(\id_\cL\otimes\proj{\Psi}_{R^p})\otimes\proj{0}_{\countsys^p}$ onto the ground space of $H_0^p$ also has the required form.  Identity~\eqref{eq:effectivehamiltonianexpr} therefore follows from Theorem~\ref{thm:maingadget} and  Proposition~\ref{prop:specialrep}.

\subsubsection{Generating a vertex-term}
Consider a vertex~$v$. We construct a Hamiltonian $H^v_0$ and a perturbation $V^v$ which generates a term proportional to the vertex operator~$A(v)$. Clearly, we could use the same procedure as for plaquettes. Since~$v$ has three incident edges $\{e_0,e_1,e_2\}$, this would give the vertex operator $A(v)$ in 3rd order perturbation theory. However, this is not suitable for our purposes. Because our ultimate goal is to generate the full quantum double Hamiltonian $H_{QD}$, we will instead show how to obtain both plaquette- and vertex-operators in the same order in perturbation theory, with identical constants.

As before, we introduce auxiliary systems $\cH_{R^v}\cong\mathbb{C}^{|G|}$ and $\cH_{I^v}\cong \mathbb{C}^6$ associated with the vertex~$v$. Let 
\begin{align*}
M_i^v=M(e_i,v)=\sum_g L^g(e_i,v)_\cL\otimes (T^g_{+})_{R^v}\qquad\textrm{for } i=0,1,2
\end{align*}
and $\ket{\Psi^v}_{R^v}=\frac{1}{\sqrt{|G|}}\sum_{g\in G}\ket{g}$ be as in Proposition~\ref{prop:specialrep}.
The unperturbed Hamiltonian has the same form as~\eqref{eq:hpzerodef}, that is,
\begin{align*}
H_0^v &=-\proj{\Psi^v}_{R^v}\otimes\proj{0}_{\countsys^v}\ 
\end{align*}
and the perturbation is
\begin{align}
V^v &=\sum_{i=0}^5\left((\tilde{M}_i^v)^\dagger\otimes \ket{i+1}\bra{i}_{I^p}+h.c.\right)\qquad\textrm{ where }\label{eq:vvdef}\\
\tilde{M}_i^v &=\begin{cases} M_i^v \qquad&\textrm{ for }i=0,1,2\\
\id_{\cL R^v}\qquad&\textrm{otherwise}\ .\nonumber
\end{cases}
\end{align}
It is straightforward to prove that the operators $\{\tilde{M}_i^v\}_i$ satisfy the requirements of Theorem~\ref{thm:maingadget}, which shows (by Proposition~\ref{prop:specialrep}) that the Hamiltonian $H^v=H_0^v+\lambda V^v$ gives rise to the effective Hamiltonian
\begin{align}
\tilde{H}_{\textrm{eff}}^v(f(\lambda)) &=-2\lambda^6\cdot A(v)_\cL\otimes\proj{\Psi}_{R^v}\otimes\proj{0}_{I^v}+O(\lambda^{7})\ .
\end{align}

\subsection{Generating all plaquette/vertex-operators and $H_{QD}$\label{sec:allplaquettes}}
In the previous section, we have shown how to obtain a single vertex or plaquette-term in 6th-order perturbation theory.  We now consider the problem of generating several terms simultaneously. Our strategy is to introduce auxiliary systems $\cH_{R^v}$, $\cH_{I^v}$ for every vertex-term $A(v)$ and   $\cH_{R^p}$, $\cH_{I^p}$ for every plaquette-term~$B(p)$ we would like to generate. 
We  use the same Hamiltonians as before, that is, 
\begin{align*}
H_0^p &=-\proj{\Psi}_{R^p}\otimes\proj{0}_{\countsys^p}\qquad\qquad  H_0^v =-\proj{\Psi}_{R^v}\otimes\proj{0}_{\countsys^v}\ .
\end{align*}
Similarly, we use the operators $M_i^v,M_j^p$ and $\tilde{M}_i^v$, where the superscripts indicate the different vertices/plaquettes these operators are associated with. 

 Writing $\alpha$ and $\beta$ for arbitrary plaquettes/vertices, it is easy to see that the operators $\{M^\alpha_i\}_{\alpha,i}$ and the ground state projections $\Gamma^\alpha_0=(\id_\cL\otimes\proj{\Psi}_{R^\alpha})\otimes\proj{0}_{I^\alpha}$ obey the rules~\eqref{it:multicommfirst}--\eqref{it:multicommfourth} (due to the unitarity of the operators $\{M^\alpha_i\}_{\alpha,i}$) and the commutation relations~\eqref{eq:gammamultdef}--\eqref{eq:specialcommutativity} (because operators
 with distinct indices $\alpha\neq\beta$ act non-trivially on distinct systems).  To apply Theorem~\ref{thm:maingadget}$^\prime$, we need to consider the commutation relations between different operators $M^\alpha_i$: According to Lemma~\ref{lem:miadditionalproperty}, we only need to take care of pairs of vertex- and plaquette-terms  when the vertex is on the boundary of the plaquette. These do not commute if they act on the same edge (cf. Figure~\ref{fig:edgesimple}). In particular, we can apply  Theorem~\ref{thm:maingadget}$^\prime$ in two different ways, giving the following statements:

\subsubsection{Generating the plaquette-part of $H_{QD}$}
Consider the perturbations $V^p$  associated to plaquette $p$ defined by~\eqref{eq:vpdef}. The effective Hamiltonian corresponding to $H=\sum_{p} H_0^p+\lambda \sum_{p} V^p$ is given by
\begin{align*}
\tilde{H}_{\textrm{eff}}(f(\lambda))= -2\lambda^{6}\cdot \left(\sum_{p} B(p)\right)\otimes\left(\bigotimes_{p} \proj{\Psi}_{R^p}\otimes\proj{0}_{I^p}\right)+O(\lambda^7)\ .
\end{align*}
for some function $f(\lambda)$. An analogous statement holds for the vertex-terms.

\subsubsection{Generating the full quantum double Hamiltonian $H_{QD}$}
For every vertex $v$, let
\begin{align*}
V^v &=\sum_{i=0} ^2\left( (M^v_i)^\dagger\otimes\ket{i+1}\bra{i}_{I^v}\otimes\proj{0}_{I^{p_{-}(e_i^v)}}\otimes\proj{0}_{I^{p_{+}(e_i^v)}}+h.c.\right)+\sum_{i=3} ^5\left( \id\otimes\ket{i+1}\bra{i}_{I^v}+h.c.\right)\ ,
\end{align*}
where $p_{-}(e_i^v)$ and $p_{+}(e_i^v)$ are the plaquettes separated by the edge $e_i^v$ (this is the edge that $M^v_i$ acts on). We also define
\begin{align*}
V^p &=\sum_{i=0} ^5\left( (M^p_i)^\dagger\otimes\ket{i+1}\bra{i}_{I^p}\otimes\proj{0}_{I^{v_{-}(e_i^p)}}\otimes\proj{0}_{I^{v_{+}(e_i^p)}}+h.c.\right)\ ,
\end{align*}
where $v_{-}(e_i^p)$ and $v_{+}(e_i^p)$ are the endpoints of the edge~$e_i^p$.
The support of these operators is visualized in~\eqref{eq:visualizationsecond}, where we have omitted a diagram of the form
\psset{unit=0.8cm}
$\raisebox{-0.64cm}{\begin{pspicture}(-0.977,-0.977)(0.977,0.817)
\SpecialCoor
\rput(0.5774;30){\myhexagonsecond}
\rput(0.5774;270){\myhexagonsecond}
\rput(0.5774;150){\myhexagonsecond}
\psset{linecolor=black,linewidth=2pt}
\qdisk(0,0){2.5pt}
\end{pspicture}}$
corresponding to the second sum in the definition of $V^v$.
\psset{unit=0.3cm}

 Consider the Hamiltonian $H=\sum_{p}H_0^p+\sum_v H_0^v+\lambda\left(\sum_v V^v+\sum_p V^p\right)$.
 Then
\begin{align*}
\tilde{H}_{\textrm{eff}}(f(\lambda))= -2\lambda^{6}\cdot H_{QD}\otimes\left(\bigotimes_{v} \proj{\Psi}_{R^v}\otimes\proj{0}_{I^v}\right)\otimes\left(\bigotimes_{p} \proj{\Psi}_{R^p}\otimes\proj{0}_{I^p}\right) +O(\lambda^7)\ .
\end{align*}
for some function $f(\lambda)$.

\section{Conclusions\label{sec:conclusions}} 
Our results show how to obtain  Kitaev's quantum double-based lattice Hamiltonians  as the low-energy effective description of Hamiltonians made of simpler and more local terms. This is achieved with limited overhead by exploiting the relation between the Hamiltonian and closed anyonic (Wilson-)loops.  We believe that our techniques may  extend to  systems  such as Levin and Wen's string-net models, where the resulting reduction in complexity may be more pronounced. A major open problem concerns the convergence of the perturbation series: The current analysis only guarantees convergence for a coupling strength that scales with the system size.

\nonumsection{Acknowledgments}
\noindent
I thank Stephen Jordan, Alexei Kitaev, David Poulin and John Preskill for discussions, Ben Reichardt for comments on an earlier draft, and Todd Brun for suggesting to use energy penalties between different clocks. I also thank the anonymous referees for their suggestions. Support by NSF Grants PHY-0456720, PHY-0803371 and SNF PA00P2-126220 is gratefully acknowledged.

\nonumsection{References}

\section*{Appendix: Proof of Theorem~\ref{thm:maingadget}$^\prime$\label{sec:maingadgetvarproof}}
We follow the steps used in the proof of Theorem~\ref{thm:maingadget}. 
Again, we consider the shifted Hamiltonian
\begin{align*}
H_0 &=\sum_{\alpha}\left(\id- \Gamma_0^\alpha
\otimes\proj{0}_{\alpha}\right)\ 
\end{align*}
which has vanishing ground state energy.
As in the proof of Theorem~\ref{thm:maingadget}, we introduce operators
\begin{align*}
(W_i^\alpha)^\dagger &=(M_i^\alpha)^\dagger\otimes\ket{i+1}\bra{i}_{\alpha}\otimes \cK(\alpha,i)\ .
\end{align*}
such that the perturbation takes the form $V=\sum_{\alpha,i} ((W_i^\alpha)^\dagger+W_i^\alpha)$.  For every fixed $\alpha$, we index the operators $\cup_i\{W_{i}^\alpha,(W_i^\alpha)^\dagger\}=\{Z^\alpha_\tau\}_\tau$ by a parameter $\tau$ and sometimes also write 
$V =\sum_{\alpha,\tau} Z^\alpha_{\tau}$
where $Z^\alpha_{\tau}\in \cup_i\{W_{i}^\alpha,(W_i^\alpha)^\dagger\}$.

 First observe that by~\eqref{eq:gammamultdef}, the projections
$P_0^\alpha= \Gamma_0^\alpha\otimes \proj{0}_{\alpha}$
onto the ground space of the $\alpha$-th term in the Hamiltonian commute with one another. The projection onto the ground space of $H_0$ is given by
$P_0 =\prod_{\alpha} P^\alpha_0$. More generally, defining $P_1^\alpha=\id-P_0^\alpha$, the projection onto the eigenspace corresponding to (integer) energy $E$ is given by
\begin{align}\label{eq:energyproj}
P_E &= \sum_{\substack{(\varepsilon^{1},\ldots,\varepsilon^{L})\in\{0,1\}^L\\
\sum_\alpha \varepsilon^\alpha=E}}\prod_{\alpha} P^\alpha_{\varepsilon^\alpha}\qquad E=0,1,\ldots,L\ .
\end{align}
The reduced resolvent~\eqref{eq:reducedresolvent} is
\begin{align*}
S^\ell &=\sum_{E\neq 0}(-E)^{-\ell} P_E\qquad\textrm{ for }\qquad \ell>0\ .
\end{align*}
A typical term in $\cA^{(m)}$ (cf.~\eqref{eq:amdefinition}) takes the form
\begin{align}
P_0 VS^{\ell_1}V\cdots VS^{\ell_{m-1}}VP_0 &=\sum_{E_1,\ldots, E_{m-1}} h_\ell(E_1,\ldots, E_{m-1}) P_0 V P_{E_1}V\cdots VP_{E_{m-1}}V P_0\ \label{eq:hdefinitionb}
\end{align}
for the function $h_\ell=h_{\ell_1,\ldots,\ell_{m-1}}$ given by
\begin{align}
h_{\ell}(E_1,\ldots,E_{m-1})=\left(\prod_{i:\ell_i=0}(-1)^{\ell_i}\delta_{E_i,0}\right) \left(\prod_{i:\ell_i\neq 0} (-E_i)^{-\ell_i}\right)\ .\label{eq:helldef}
\end{align}
We expand the operators in this sum further using~\eqref{eq:energyproj}, getting
\begin{align}
 P_0 V P_{E_1}V\cdots VP_{E_{m-1}}V P_0&=\sum_{\varepsilon=(\varepsilon_j^\alpha)^{\alpha=1,\ldots,L}_{j=1,\ldots,m-1}}P_0 V P(\varepsilon_1)V\cdots VP(\varepsilon_{m-1})VP_0\qquad\textrm{ with }\label{eq:prodc}\\
P(\varepsilon_j)&=P(\varepsilon_j^1,\ldots,\varepsilon_j^L):=\prod_{\alpha} P^\alpha_{\varepsilon^\alpha_j}\ ,\nonumber
\end{align}
where the sum is over all matrices $\varepsilon$ with entries in $\{0,1\}$ satisfying $\sum_{\alpha}\varepsilon_j^\alpha =E_j$ for all $j=1,\ldots,m-1$. Combining~\eqref{eq:amdefinition},~\eqref{eq:hdefinitionb} and \eqref{eq:prodc}, we obtain
\begin{align}\label{eq:aresummed}
\cA^{(m)}&=\sum_{\ell}\sum_{\varepsilon} h_{\ell}\left(\sum_{\alpha}\varepsilon^\alpha_1,\ldots,\sum_{\alpha}\varepsilon^\alpha_{m-1}\right)P_0 V P(\varepsilon_1)V\cdots VP(\varepsilon_{m-1})VP_0\ ,
\end{align}
where the sums are over all $\ell=(\ell_1,\ldots,\ell_{m-1})\in \cP_{m-1}$ and all matrices $\varepsilon=(\varepsilon_j^\alpha)^{\alpha=1,\ldots,L}_{j=1,\ldots,m-1}$ with entries in $\{0,1\}$.

Inserting the decomposition of the perturbation $V$ as a sum of operators $Z^{\alpha}_\tau$, the operators in~\eqref{eq:aresummed} are
\begin{align}
P_0VP(\varepsilon_1)\cdots P(\varepsilon_{m-1})VP_0 &=\sum_{\substack{(\alpha_1,\ldots,\alpha_m)\\
(\tau_1,\ldots,\tau_m)}} P_0 Z^{\alpha_1}_{\tau_1}P(\varepsilon_1)\cdots P(\varepsilon_{m-1}) Z^{\alpha_m}_{\tau_m}P_0\ . \label{eq:sumpz}
\end{align}
Consider a term $P_0 Z^{\alpha_1}_{\tau_1}P(\varepsilon_1)\cdots P(\varepsilon_{m-1}) Z^{\alpha_m}_{\tau_m}P_0$.  Observe that the $j$-th operator $Z$ only acts non-trivially  on $\cH\otimes\cH_{I^{\alpha_j}}$. Defining
 ${\bf Y}=(Y_j^\alpha)_j^{\alpha}$ as 
\begin{align*}
Y^\alpha_j&=\begin{cases}
Z^{\alpha_j}_{\tau_j}\qquad&\textrm{ if }\alpha=\alpha_j\\
\id &\textrm{otherwise}\ ,
\end{cases}
\end{align*}
we can write $Z^{\alpha_j}_{\tau_j}=\prod_{\alpha} Y^\alpha_j=:Y_j$, 
or
\begin{align}
P_0 Z^{\alpha_1}_{\tau_1}P(\varepsilon_1)\cdots P(\varepsilon_{m-1}) Z^{\alpha_m}_{\tau_m}P_0&=P_0 Y_1 P(\varepsilon_1)Y_2\cdots P(\varepsilon_{m-1})Y_mP_0:=\Theta(\varepsilon,{\bf Y})\ .\label{eq:thetaexpanded}
\end{align}
In particular, this allows us to rewrite~\eqref{eq:sumpz} as
\begin{align}
P_0VP(\varepsilon_1)\cdots P(\varepsilon_{m-1})VP_0 &=\sum_{\bf Y} \Theta(\varepsilon,{\bf Y})\label{eq:vpsumrestricted}
\end{align}
where the sum is over a restricted set of  matrices ${\bf Y}=(Y^\alpha_j)^{\alpha}_j$ of operators.  Combining~\eqref{eq:aresummed} with~\eqref{eq:vpsumrestricted} gives the following generalization of Lemma~\ref{lem:Ausefulform}.
\newtheorem*{varlemmauseful}{Lemma~\ref{lem:Ausefulform}$^\prime$}
\vspace*{12pt}
\noindent
\begin{varlemmauseful}
The operator $\cA^{(m)}$ is a linear combination
\begin{align*}
\cA^{(m)}&=\sum_{\varepsilon,{\bf Y}}g_m(\varepsilon)\Theta(\varepsilon,{\bf Y})\ ,
\end{align*}
where the sum is over all matrices $\varepsilon=\{\varepsilon_j^{\alpha}\}^{\alpha=1,\ldots,L}_{j=1,\ldots,m-1}$ with entries in~$\{0,1\}$, and all ${\bf Y}=(Y_j^\alpha)_j^{\alpha}$ with the property that for every $j$, there is exactly one $\alpha_j$ such that
\begin{align*}
Y_j^{\alpha_j}&\in \cup_i\{W^{\alpha_j}_i,(W^{\alpha_j}_i)^\dagger \}\\
Y_j^\alpha&=\id\qquad\textrm{ for }\alpha\neq\alpha_j\ .
\end{align*}
The operators $\Theta(\varepsilon,{\bf Y})$ are defined by~\eqref{eq:thetaexpanded}. The function $g_m$ is given by 
\begin{align*}
g_m(\varepsilon ) =\sum_{\ell\in \cP_{m-1}}h_\ell\left(\sum_\alpha \varepsilon^\alpha_1,\ldots,\sum_\alpha \varepsilon^\alpha_{m-1}\right)\ ,
\end{align*} where $h_\ell$ is defined by~\eqref{eq:helldef}.
\end{varlemmauseful}
\vspace*{12pt}
\noindent
As in the proof of Theorem~\ref{thm:maingadget},  we introduce a diagrammatic notation for the operators~$\Theta(\varepsilon,{\bf Y})$. In essence, we stack the diagrams of Section~\ref{sec:perturbativegadgetsordered}, introducing in addition an appropriately placed  horizontal arrow  when $Y_j^\alpha=\id$.  An example is
\begin{align}
\Theta(\varepsilon,{\bf Y})&=\qquad 
\begin{matrix}\begin{pspicture}(0,-1)(4,1)
\psgrid[subgriddiv=1,linecolor=lightgray,griddots=5,gridlabels=5pt](0,-1)(4,1)
\rput(0,0){\zzerobig}\rput(4,0){\zzerobig}
\rput(1,0){\zone}\rput(2,0){\zone}\rput(3,0){\zone}
\rput(0,0){\zdown}
\rput(1,1){\zleft}
\rput(2,1){\zleft}
\rput(3,1){\zup}
\end{pspicture}\\
\\
\begin{pspicture}(0,-1)(4,1)
\psgrid[subgriddiv=1,linecolor=lightgray,griddots=5,gridlabels=5pt](0,-1)(4,1)
\rput(0,0){\zzerobig}\rput(4,0){\zzerobig}
\rput(1,0){\zzero}\rput(2,0){\zzero}\rput(3,0){\zzero}
\rput(0,0){\zleft}
\rput(1,0){\zleft}
\rput(2,0){\zleft}
\rput(3,0){\zleft}
\end{pspicture}\\
\\
\begin{pspicture}(0,-1)(4,1)
\psgrid[subgriddiv=1,linecolor=lightgray,griddots=5,gridlabels=5pt](0,-1)(4,1)
\rput(0,0){\zzerobig}\rput(4,0){\zzerobig}
\rput(1,0){\zzero}\rput(2,0){\zone}\rput(3,0){\zzero}
\rput(0,0){\zleft}
\rput(1,0){\zup}
\rput(2,-1){\zdown}
\rput(3,0){\zleft}
\end{pspicture}
\end{matrix}\label{eq:thetaepsY}
\qquad\textrm{ for }
\qquad
\varepsilon =\left(\begin{matrix}
1&1&1\\
0&0&0\\
0&1&0
\end{matrix}\right)\qquad{\bf Y}=\left(\begin{matrix}
W^{1}_0&\id&\id & (W^{1}_0)^\dagger\\
\id&\id&\id&\id\\
\id& (W_{n-1}^3)^\dagger & W_{n-1}^3 &\id 
\end{matrix}\right)\ .
\end{align}
The corresponding operator can be read off  as 
\begin{align*}
\Theta(\varepsilon,{\bf Y})&=P_0W_0^{1}(P_1^{1}P_0^2P_0^3)(W_{n-1}^3)^\dagger (P_1^{1}P_0^2P_1^3)W_{n-1}^3(P_1^{1}P_0^2P_0^3)(W_0^{1})^\dagger P_0\ .
\end{align*}
Note that we are only interested in a subset of operators $\Theta(\varepsilon,{\bf Y})$, as specified by the condition on ${\bf Y}$ in Lemma~\ref{lem:Ausefulform}$^\prime$. We will express this as a rule; we demand that
\begin{enumerate}\setcounter{enumi}{-1}
\item\label{it:zerocond}
On each vertical, there is exactly one non-horizontal arrow. 
\end{enumerate}
From now on, we will only consider diagrams obeying this rule.  In fact, we will further restrict the set of diagrams we study. As before, only certain diagrams correspond to non-zero operators $\Theta(\varepsilon,{\bf Y})$, as can be seen by inspecting the definitions. In particular,
the previously established rules apply to every ``subdiagram'' defined by
$\left(\varepsilon^\alpha=(\varepsilon^\alpha_1,\ldots,\varepsilon^\alpha_m),{\bf Y}^\alpha=(Y^\alpha_1,\ldots,Y^\alpha_m)\right)$. That is, every diagram corresponding to a non-zero operator $\Theta(\varepsilon,{\bf Y})$ satisfies for all $\alpha$
\begin{enumerate}
\item\label{it:firstcond}
the sequence of arrows defined by ${\bf Y}^\alpha$ defines a continuous path which
\item
starts at $(m,0)$ (modulo $(0,n)$)
\item
ends at $(0,0)$ (modulo $(0,n)$)
\item
goes through the points $(\varepsilon^\alpha_i,0)$ (modulo $(0,n)$) for all $i$ with $\varepsilon^\alpha_i=0$.\label{it:fourthcond}
\end{enumerate}
In addition to rules~\eqref{it:zerocond}--\eqref{it:fourthcond}, diagrams corresponding to non-zero operators $\Theta(\varepsilon,{\bf Y})$ satisfy further conditions as a result of the horizontal arrows and the operators $\cK(\alpha,i)$. In particular,  if the circles on the verticals defined by the endpoints of a horizontal arrow are of a different color, then the corresponding operator vanishes. In other words, we have the additional rule
\begin{enumerate}\setcounter{enumi}{4}
\item
for every horizontal arrow (corresponding to $Y^{\alpha}_i=\id$), the circles
at the horizontal positions defined by its endpoints must have the same color ($\varepsilon^\alpha_{i-1}=\varepsilon^{\alpha}_i$).
\end{enumerate}
 For example, any diagram containing
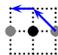
\begin{align*}
\raisebox{-0.25cm}{\begin{pspicture}(0,-1)(2,1)
\psgrid[subgriddiv=1,linecolor=lightgray,griddots=5,gridlabels=0pt](0,-1)(2,1)
\rput(0,0){\zzero}
\rput(1,0){\zone}
\rput(2,0){\zzero}
\rput(0,1){\zleft}
\rput(1,1){\zup}
\end{pspicture}}
\end{align*}
gives $\Theta(\varepsilon,{\bf Y})=0$ because this operator contains a product of the form $P^\alpha_0Y^\beta_jP^\alpha_1=P_0^\alpha P_1^\alpha Y_j^\beta=0$ (here we used~\eqref{eq:secondcommutat}). Now consider the effect of the operators $\cK(\alpha,i)$. They imply that 
applying $W^\alpha_{j}$ (or its adjoint) from the left to an operator can only lead to a non-zero result if the operator is not killed by the projections $\proj{0}_\beta$ for all $\beta\in\chi(\alpha,j)$. 
This translates into the rule
\begin{enumerate}\setcounter{enumi}{5}
\item\label{it:lastrule}
For every diagonal arrow  in a subdiagram $\alpha$ (i.e., $Y^{\alpha}_i\neq\id$), the following holds: Every horizontal arrow (i.e.,~$Y^{\beta}_i$) corresponding to a subdiagram $\beta\in\chi(\alpha,i)$ in the same vertical lies on the horizontal axis.
\end{enumerate}
For example,~\eqref{eq:thetaepsY} satisfies this rule if and only if $1\not\in\chi(3,1)$.

We will call a diagram obeying rules~\eqref{it:zerocond}--\eqref{it:lastrule} valid and restrict our attention to such diagrams. Again, we can write the operators $\Theta(\varepsilon,{\bf Y})$ associated with valid diagrams in terms of the operators $M_i$, $\Gamma_i$. This is done by substituting~$W$ by~$M$ and every  circle that is touched by~$\Gamma_0$  or~$\Gamma_1$ depending on its color. For our example, we get (assuming that the diagram is valid)
\begin{align*}
\Theta(\varepsilon,{\bf Y})&=\Gamma M_0^{1}(\Gamma_0^2\Gamma_0^3)(M_{n-1}^3)^\dagger \Gamma_0^2M_{n-1}^3(\Gamma_0^2\Gamma_0^3)(M_0^{1})^\dagger\Gamma\otimes\proj{0}^{\otimes 3}\ .
\end{align*}

We now prove a few substitution rules which we will use to simplify diagrams.
These affect two systems $(\alpha)$ and $(\beta)$ (and leave the others invariant). To state the rules, we only depict the relevant parts of the subdiagrams 
$(\varepsilon^\alpha,{\bf Y}^\alpha)$ and $(\varepsilon^\beta,{\bf Y}^\beta)$. The colored circles in the following diagrams may be arbitrary.
\vspace*{12pt}
\noindent
\begin{lemma}\label{lem:substitution}
In every valid diagram, we have the following substitution rules.
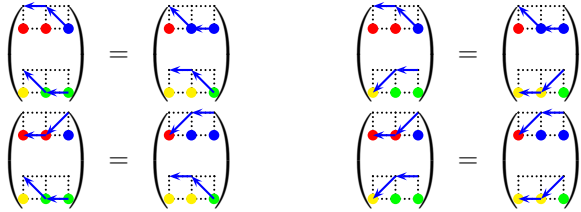
\begin{enumerate}[(a)]
\item\label{it:firstcomm}
\begin{align*}
\left(\begin{matrix}
\begin{pspicture}(0,0)(2,1)
\psgrid[subgriddiv=1,linecolor=lightgray,griddots=5,gridlabels=0pt](0,0)(2,1)
\rput(0,0){\zundefined}\rput(1,0){\zundefined}\rput(2,0){\zundefinedsecond}
\rput(0,1){\zleft}
\rput(1,1){\zup}
\end{pspicture}\\
\\
\begin{pspicture}(0,0)(2,1)
\psgrid[subgriddiv=1,linecolor=lightgray,griddots=5,gridlabels=0pt](0,0)(2,1)
\rput(0,0){\zundefinedfourth}\rput(1,0){\zundefinedthird}\rput(2,0){\zundefinedthird}
\rput(1,0){\zleft}
\rput(0,1){\zup}
\end{pspicture}
\end{matrix}\right)\ &=\ 
\left(\begin{matrix}
\begin{pspicture}(0,0)(2,1)
\psgrid[subgriddiv=1,linecolor=lightgray,griddots=5,gridlabels=0pt](0,0)(2,1)
\rput(0,0){\zundefined}\rput(1,0){\zundefinedsecond}\rput(2,0){\zundefinedsecond}
\rput(1,0){\zleft}
\rput(0,1){\zup}
\end{pspicture}\\
\\
\begin{pspicture}(0,0)(2,1)
\psgrid[subgriddiv=1,linecolor=lightgray,griddots=5,gridlabels=0pt](0,0)(2,1)
\rput(0,0){\zundefinedfourth}\rput(1,0){\zundefinedfourth}\rput(2,0){\zundefinedthird}
\rput(0,1){\zleft}
\rput(1,1){\zup}
\end{pspicture}
\end{matrix}\right)\qquad\qquad
\left(\begin{matrix}
\begin{pspicture}(0,0)(2,1)
\psgrid[subgriddiv=1,linecolor=lightgray,griddots=5,gridlabels=0pt](0,0)(2,1)
\rput(0,0){\zundefined}\rput(1,0){\zundefined}\rput(2,0){\zundefinedsecond}
\rput(0,1){\zleft}
\rput(1,1){\zup}
\end{pspicture}\\
\\
\begin{pspicture}(0,0)(2,1)
\psgrid[subgriddiv=1,linecolor=lightgray,griddots=5,gridlabels=0pt](0,0)(2,1)
\rput(0,0){\zundefinedfourth}\rput(1,0){\zundefinedthird}\rput(2,0){\zundefinedthird}
\rput(0,0){\zdown}
\rput(1,1){\zleft}
\end{pspicture}
\end{matrix}\right)\ =\ 
\left(\begin{matrix}
\begin{pspicture}(0,0)(2,1)
\psgrid[subgriddiv=1,linecolor=lightgray,griddots=5,gridlabels=0pt](0,0)(2,1)
\rput(0,0){\zundefined}\rput(1,0){\zundefinedsecond}\rput(2,0){\zundefinedsecond}
\rput(1,0){\zleft}
\rput(0,1){\zup}
\end{pspicture}\\
\\
\begin{pspicture}(0,0)(2,1)
\psgrid[subgriddiv=1,linecolor=lightgray,griddots=5,gridlabels=0pt](0,0)(2,1)
\rput(0,0){\zundefinedfourth}\rput(1,0){\zundefinedfourth}\rput(2,0){\zundefinedthird}
\rput(0,0){\zleft}
\rput(1,0){\zdown}
\end{pspicture}
\end{matrix}\right)\\ 
\left(\begin{matrix}
\begin{pspicture}(0,0)(2,1)
\psgrid[subgriddiv=1,linecolor=lightgray,griddots=5,gridlabels=0pt](0,0)(2,1)
\rput(0,0){\zundefined}\rput(1,0){\zundefined}\rput(2,0){\zundefinedsecond}
\rput(1,0){\zdown}
\rput(0,0){\zleft}
\end{pspicture}\\
\\
\begin{pspicture}(0,0)(2,1)
\psgrid[subgriddiv=1,linecolor=lightgray,griddots=5,gridlabels=0pt](0,0)(2,1)
\rput(0,0){\zundefinedfourth}\rput(1,0){\zundefinedthird}\rput(2,0){\zundefinedthird}
\rput(1,0){\zleft}
\rput(0,1){\zup}
\end{pspicture}
\end{matrix}\right)\ &=\ 
\left(\begin{matrix}
\begin{pspicture}(0,0)(2,1)
\psgrid[subgriddiv=1,linecolor=lightgray,griddots=5,gridlabels=0pt](0,0)(2,1)
\rput(0,0){\zundefined}\rput(1,0){\zundefinedsecond}\rput(2,0){\zundefinedsecond}
\rput(0,0){\zdown}
\rput(1,1){\zleft}
\end{pspicture}\\
\\
\begin{pspicture}(0,0)(2,1)
\psgrid[subgriddiv=1,linecolor=lightgray,griddots=5,gridlabels=0pt](0,0)(2,1)
\rput(0,0){\zundefinedfourth}\rput(1,0){\zundefinedfourth}\rput(2,0){\zundefinedthird}
\rput(0,1){\zleft}
\rput(1,1){\zup}
\end{pspicture}
\end{matrix}\right)\qquad\qquad
\left(\begin{matrix}
\begin{pspicture}(0,0)(2,1)
\psgrid[subgriddiv=1,linecolor=lightgray,griddots=5,gridlabels=0pt](0,0)(2,1)
\rput(0,0){\zundefined}\rput(1,0){\zundefined}\rput(2,0){\zundefinedsecond}
\rput(0,0){\zleft}
\rput(1,0){\zdown}
\end{pspicture}\\
\\
\begin{pspicture}(0,0)(2,1)
\psgrid[subgriddiv=1,linecolor=lightgray,griddots=5,gridlabels=0pt](0,0)(2,1)
\rput(0,0){\zundefinedfourth}\rput(1,0){\zundefinedthird}\rput(2,0){\zundefinedthird}
\rput(1,1){\zleft}
\rput(0,0){\zdown}
\end{pspicture}
\end{matrix}\right)\ =\ 
\left(\begin{matrix}
\begin{pspicture}(0,0)(2,1)
\psgrid[subgriddiv=1,linecolor=lightgray,griddots=5,gridlabels=0pt](0,0)(2,1)
\rput(0,0){\zundefined}\rput(1,0){\zundefinedsecond}\rput(2,0){\zundefinedsecond}
\rput(0,0){\zdown}
\rput(1,1){\zleft}
\end{pspicture}\\
\\
\begin{pspicture}(0,0)(2,1)
\psgrid[subgriddiv=1,linecolor=lightgray,griddots=5,gridlabels=0pt](0,0)(2,1)
\rput(0,0){\zundefinedfourth}\rput(1,0){\zundefinedfourth}\rput(2,0){\zundefinedthird}
\rput(0,0){\zleft}
\rput(1,0){\zdown}
\end{pspicture}
\end{matrix}\right)
\end{align*}
where we assume that the endpoint of the upper sequence of arrows is not on the horizontal axis (apart from that, the vertical/horizontal positions of the arrows may be arbitrary).

\item\label{it:secondcomm}
Arrows can be commuted past triangles situated on the horizontal axis,  that is,\newcommand*{\triangleleftzz}
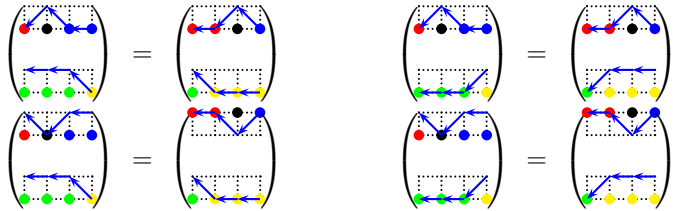
{\begin{pspicture}(0,0)(3,1)
\psgrid[subgriddiv=1,linecolor=lightgray,griddots=5,gridlabels=0pt](0,0)(3,1)
\rput(0,0){\zundefined}\rput(3,0){\zundefinedsecond}\rput(2,0){\zundefinedsecond}\rput(1,0){\zone}
\rput(0,0){\zdown}
\rput(1,1){\zup}
\rput(2,0){\zleft}
\end{pspicture}}

\newcommand*{\btriangleleftzz}{\begin{pspicture}(0,0)(3,1)
\psgrid[subgriddiv=1,linecolor=lightgray,griddots=5,gridlabels=0pt](0,0)(3,1)
\rput(0,0){\zundefined}\rput(3,0){\zundefinedsecond}\rput(2,0){\zundefinedsecond}\rput(1,0){\zone}
\rput(1,0){\zdown}
\rput(0,1){\zup}
\rput(2,1){\zleft}
\end{pspicture}}

\newcommand*{\trianglerightzz}{\begin{pspicture}(0,0)(3,1)
\psgrid[subgriddiv=1,linecolor=lightgray,griddots=5,gridlabels=0pt](0,0)(3,1)
\rput(0,0){\zundefined}
\rput(1,0){\zundefined}
\rput(3,0){\zundefinedsecond}\rput(2,0){\zone}
\rput(2,1){\zup}
\rput(1,0){\zdown}
\rput(0,0){\zleft}
\end{pspicture}}

\newcommand*{\btrianglerightzz}{\begin{pspicture}(0,0)(3,1)
\psgrid[subgriddiv=1,linecolor=lightgray,griddots=5,gridlabels=0pt](0,0)(3,1)
\rput(0,1){\zundefined}\rput(3,1){\zundefinedsecond}\rput(1,1){\zundefined}\rput(2,1){\zone}
\rput(1,1){\zup}
\rput(0,1){\zleft}
\rput(2,0){\zdown}
\end{pspicture}}

\newcommand*{\uparrowrightzz}{
\begin{pspicture}(0,0)(3,1)
\psgrid[subgriddiv=1,linecolor=lightgray,griddots=5,gridlabels=0pt](0,0)(3,1)
\rput(0,0){\zundefinedthird}\rput(1,0){\zundefinedthird}\rput(2,0){\zundefinedthird}\rput(3,0){\zundefinedfourth}
\rput(0,1){\zleft}
\rput(1,1){\zleft}
\rput(2,1){\zup}
\end{pspicture}
}

\newcommand*{\downarrowrightzz}{
\begin{pspicture}(0,0)(3,1)
\psgrid[subgriddiv=1,linecolor=lightgray,griddots=5,gridlabels=0pt](0,0)(3,1)
\rput(0,0){\zundefinedthird}\rput(1,0){\zundefinedthird}\rput(2,0){\zundefinedthird}\rput(3,0){\zundefinedfourth}
\rput(0,0){\zleft}
\rput(1,0){\zleft}
\rput(2,0){\zdown}
\end{pspicture}
}

\newcommand*{\uparrowleftzz}{
\begin{pspicture}(0,0)(3,1)
\psgrid[subgriddiv=1,linecolor=lightgray,griddots=5,gridlabels=0pt](0,0)(3,1)
\rput(0,0){\zundefinedthird}\rput(1,0){\zundefinedfourth}\rput(2,0){\zundefinedfourth}\rput(3,0){\zundefinedfourth}
\rput(0,1){\zup}
\rput(1,0){\zleft}
\rput(2,0){\zleft}
\end{pspicture}
}

\newcommand*{\downarrowleftzz}{
\begin{pspicture}(0,0)(3,1)
\psgrid[subgriddiv=1,linecolor=lightgray,griddots=5,gridlabels=0pt](0,0)(3,1)
\rput(0,0){\zundefinedthird}\rput(1,0){\zundefinedfourth}\rput(2,0){\zundefinedfourth}\rput(3,0){\zundefinedfourth}
\rput(0,0){\zdown}
\rput(1,1){\zleft}
\rput(2,1){\zleft}
\end{pspicture}
}

\begin{align*}
\left(\begin{matrix}
\triangleleftzz\\
\\
\uparrowrightzz
\end{matrix}\right)&\ =\ 
\left(\begin{matrix}
\trianglerightzz\\
\\
\uparrowleftzz
\end{matrix}\right)\qquad\qquad 
\left(\begin{matrix}
\triangleleftzz\\
\\
\downarrowrightzz
\end{matrix}\right)\ =\ 
\left(\begin{matrix}
\trianglerightzz\\
\\
\downarrowleftzz
\end{matrix}\right)\\ 
\left(\begin{matrix}
\btriangleleftzz\\
\\
\uparrowrightzz
\end{matrix}\right)&\ =\ 
\left(\begin{matrix}
\btrianglerightzz\\
\\
\uparrowleftzz
\end{matrix}\right)\qquad\qquad 
\left(\begin{matrix}
\btriangleleftzz\\
\\
\downarrowrightzz
\end{matrix}\right)\ =\ 
\left(\begin{matrix}
\btrianglerightzz\\
\\
\downarrowleftzz
\end{matrix}\right)
\end{align*}
where we assume that the endpoint of the upper sequence of arrows is on the horizontal axis.
\end{enumerate}
\end{lemma}
\vspace*{12pt}
\noindent
{\bf Proof: }We show the first rule in~\eqref{it:firstcomm}.  Let $\alpha$ ($\beta$) be the index corresponding to the upper (lower) sequence of arrows. 
Because of the commutation relations~\eqref{eq:gammamultdef}--\eqref{eq:specialcommutativity}, we can restrict our attention to operators with these indices. The diagram on the left then corresponds to an operator of the form
\[
(\Lambda^\alpha_L\Lambda^\beta_L)(M_j^\beta)^\dagger (\Lambda_{L}^\alpha\Lambda_{R}^\beta)(M_i^\alpha)^\dagger(\Lambda^\alpha_R\Lambda^\beta_R)\ ,
\]
where $\Lambda^\alpha_R,\Lambda^\alpha_L\in\{\id, \Gamma_0^\alpha,\Gamma_1^\alpha\}$ and similarly for $\beta$. Since we assume $i\neq n-1$ and that this is a valid diagram, we have $\alpha\not\in\chi(\beta,j)$, that is, $[M_j^\beta,M_i^\alpha]=0$. Therefore
\begin{align*}
(M_j^\beta)^\dagger\Lambda_{L}^\alpha\Lambda_{R}^\beta(M_i^\alpha)^\dagger
&=\Lambda_{L}^\alpha(M_j^\beta)^\dagger(M_i^\alpha)^\dagger\Lambda_{R}^\beta\\
&=\Lambda_{L}^\alpha(M_i^\alpha)^\dagger(M_j^\beta)^\dagger\Lambda_{R}^\beta
\end{align*}
where we used~\eqref{eq:secondcommutat} twice to obtain the first identity and the commutativity of $M_j^\beta$ and $M_i^\alpha$. Multiplying this identity from the right with $\Lambda^\alpha_R\Lambda^\beta_R$ and from the left  with  $\Lambda^\alpha_L\Lambda^\beta_L$ gives
\begin{align*}
(\Lambda^\alpha_L\Lambda^\beta_L)(M_j^\beta)^\dagger (\Lambda_{L}^\alpha\Lambda_{R}^\beta)(M_i^\alpha)^\dagger(\Lambda^\alpha_R\Lambda^\beta_R)&=
(\Lambda^\alpha_L\Lambda^\beta_L)\Lambda_{L}^\alpha(M_i^\alpha)^\dagger(M_j^\beta)^\dagger\Lambda_{R}^\beta
(\Lambda^\alpha_R\Lambda^\beta_R)\\
&=(\Lambda^\alpha_L\Lambda^\beta_L)\Lambda^\beta_L(M_i^\alpha)^\dagger(M_j^\beta)^\dagger\Lambda^\alpha_R(\Lambda^\alpha_R\Lambda^\beta_R)\\
&=(\Lambda^\alpha_L\Lambda^\beta_L)(M_i^\alpha)^\dagger(\Lambda^\alpha_R\Lambda^\beta_L)(M_j^\beta)^\dagger(\Lambda^\alpha_R\Lambda^\beta_R)
\end{align*}
where we used the fact that the operators $\Lambda^{(\cdot)}_{\cdot}$ are commuting projections and~\eqref{eq:secondcommutat}. This is the first statement in~\eqref{it:firstcomm}; the other claims can be derived in a similar manner.

Consider the first identity in part~\eqref{it:secondcomm} of the lemma. Note that the diagram on the left corresponds to an  operator of the form
\begin{align*}
(\Lambda^\alpha_L\Lambda^\beta_L)M^\alpha_0\Lambda^\beta_L(M^\alpha_0)^\dagger (\Lambda^\alpha_R\Lambda^\beta_L)(M^\beta_j)^\dagger(\Lambda^\alpha_R\Lambda^\beta_R)
\end{align*}
where $\Lambda^\alpha_R\in \{\Gamma^\alpha_0,\Gamma^\alpha_1\}$ and $\Lambda^\beta_R,\Lambda^\beta_L\in\{\id,\Gamma^\beta_0,\Gamma^\beta_1\}$. Again using commutativity (in particular assumption~\eqref{eq:specialcommutativity}) and the fact that the operators $\Lambda$ are projections, this can be reorganized into
\begin{align*}
(\Lambda^\alpha_L\Lambda^\beta_L) M_0^\alpha (M_0^\alpha)^\dagger (M_j^\beta)^\dagger  (\Lambda_R^\alpha\Lambda_R^\beta) &=(\Lambda^\alpha_L\Lambda^\beta_L) (M_j^\beta)^\dagger M_0^\alpha (M_0^\alpha)^\dagger   (\Lambda_R^\alpha\Lambda_R^\beta)\\
&=(\Lambda^\alpha_L\Lambda^\beta_L) (M_j^\beta)^\dagger (\Lambda^\alpha_L\Lambda^\beta_R) M_0^\alpha\Lambda^\beta_R (M_0^\alpha)^\dagger   (\Lambda_R^\alpha\Lambda_R^\beta)
\end{align*} This implies the first substitution rule in~\eqref{it:secondcomm}; again, the remaining rules are shown in a similar manner and we omit the proof. \square\,

In analogy with~\eqref{eq:updowndiagrams}, we define certain operators of order~$n$ which will give us the desired terms in the effective Hamiltonian. These diagrams contain one of the ``diagonal'' diagrams in~\eqref{eq:diagonaldiagramspicture} as a subdiagram. They are defined as follows: For every $\alpha$, we set
\begin{align}
\Theta_n^{\alpha,\downarrow}&=\Theta(\varepsilon,{\bf Y})\qquad\textrm{ where}\textrm{ for all }j\textrm{ and } \beta\neq \alpha\qquad  
\begin{matrix}
\varepsilon_j^{\alpha}=1, \varepsilon_j^\beta=0 \\
Y^{\alpha}_j=W_{j-1}^{\alpha}, Y^{\beta}_j=\id
\end{matrix}\nonumber\\
\Theta_n^{\alpha,\uparrow}&=\Theta(\varepsilon,{\bf Y})\qquad\textrm{ where}\textrm{ for all }j\textrm{ and } \beta\neq \alpha\qquad  
\begin{matrix}
\varepsilon_j^{\alpha}=1, \varepsilon_j^\beta=0 \\
Y^{\alpha}_j=(W_{n-j}^{\alpha})^\dagger, Y^{\beta}_j=\id\ .
\end{matrix}\label{eq:uparrowdiagrams}
\end{align}
Our aim will be to show that the operators
corresponding to diagrams that are not of the form~\eqref{eq:uparrowdiagrams} are proportional to the projection~$P_0$. We will obtain such a generalization of Lemma~\ref{lem:centralproportionality} by transforming the diagram in a sequence of steps into a certain form. The manipulations of Lemma~\ref{lem:substitution} are essential for this purpose.

More precisely, we will define three intermediate ``standard'' forms $S1$, $S2$ and $S3$ of diagrams. An example of the transformations we will use is
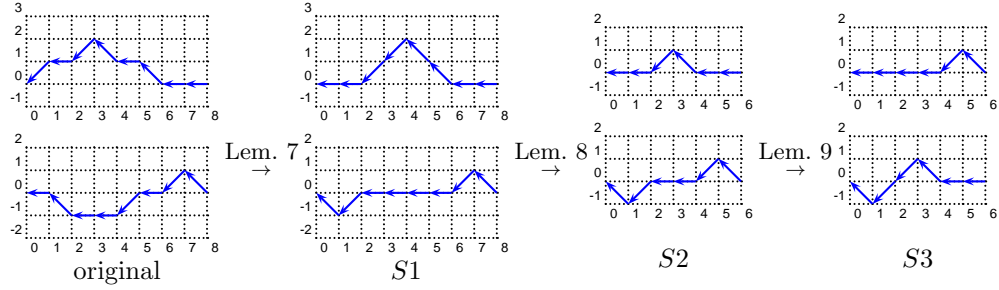
\vspace{0.05cm}
\begin{align*}
\begin{matrix}\begin{pspicture}(0,-1)(8,3)
\psgrid[subgriddiv=1,linecolor=lightgray,griddots=5,gridlabels=5pt](0,-1)(8,3)
\rput(0,0){\zdown}
\rput(1,1){\zleft}
\rput(2,1){\zdown}
\rput(3,2){\zup}
\rput(4,1){\zleft}
\rput(5,1){\zup}
\rput(6,0){\zleft}
\rput(7,0){\zleft}
\end{pspicture}\\
\\
\begin{pspicture}(0,-1)(8,2)
\psgrid[subgriddiv=1,linecolor=lightgray,griddots=5,gridlabels=5pt](0,-2)(8,2)
\rput(0,0){\zleft}
\rput(1,0){\zup}
\rput(2,-1){\zleft}
\rput(3,-1){\zleft}
\rput(4,-1){\zdown}
\rput(5,0){\zleft}
\rput(6,0){\zdown}
\rput(7,1){\zup}
\end{pspicture}\\
\\
\textrm{original} 
\end{matrix}\ \ \underset{\rightarrow}{\textrm{\small Lem.~\ref{lem:firststandardform}}}\ \ 
\begin{matrix}\begin{pspicture}(0,-1)(8,3)
\psgrid[subgriddiv=1,linecolor=lightgray,griddots=5,gridlabels=5pt](0,-1)(8,3)
\rput(0,0){\zleft}
\rput(1,0){\zleft}
\rput(2,0){\zdown}
\rput(3,1){\zdown}
\rput(4,2){\zup}
\rput(5,1){\zup}
\rput(6,0){\zleft}
\rput(7,0){\zleft}
\end{pspicture}\\
\\
\begin{pspicture}(0,-1)(8,2)
\psgrid[subgriddiv=1,linecolor=lightgray,griddots=5,gridlabels=5pt](0,-2)(8,2)
\rput(0,0){\zup}
\rput(1,-1){\zdown}
\rput(2,0){\zleft}
\rput(3,0){\zleft}
\rput(4,0){\zleft}
\rput(5,0){\zleft}
\rput(6,0){\zdown}
\rput(7,1){\zup}
\end{pspicture}\\
\\
S1
\end{matrix}\ \ \underset{\rightarrow}{\textrm{\small Lem.~\ref{lem:secondstandardform}}}\ \ \begin{matrix}\begin{pspicture}(0,-1)(6,2)
\psgrid[subgriddiv=1,linecolor=lightgray,griddots=5,gridlabels=5pt](0,-1)(6,2)
\rput(0,0){\zleft}
\rput(1,0){\zleft}
\rput(2,0){\zdown}
\rput(3,1){\zup}
\rput(4,0){\zleft}
\rput(5,0){\zleft}
\end{pspicture}\\
\\
\begin{pspicture}(0,-1)(6,2)
\psgrid[subgriddiv=1,linecolor=lightgray,griddots=5,gridlabels=5pt](0,-1)(6,2)
\rput(0,0){\zup}
\rput(1,-1){\zdown}
\rput(2,0){\zleft}
\rput(3,0){\zleft}
\rput(4,0){\zdown}
\rput(5,1){\zup}
\end{pspicture}\\
\\
S2
\end{matrix}
\ \ \underset{\rightarrow}{\textrm{\small Lem.~\ref{lem:thirdstandardform}}}\ \ \begin{matrix}\begin{pspicture}(0,-1)(6,2)
\psgrid[subgriddiv=1,linecolor=lightgray,griddots=5,gridlabels=5pt](0,-1)(6,2)
\rput(0,0){\zleft}
\rput(1,0){\zleft}
\rput(4,0){\zdown}
\rput(5,1){\zup}
\rput(2,0){\zleft}
\rput(3,0){\zleft}
\end{pspicture}\\
\\
\begin{pspicture}(0,-1)(6,2)
\psgrid[subgriddiv=1,linecolor=lightgray,griddots=5,gridlabels=5pt](0,-1)(6,2)
\rput(0,0){\zup}
\rput(1,-1){\zdown}
\rput(4,0){\zleft}
\rput(5,0){\zleft}
\rput(2,0){\zdown}
\rput(3,1){\zup}
\end{pspicture}\\
\\
S3
\end{matrix}
\end{align*}
where we indicated the lemmas explaining these transformations (we omit the circles for simplicity). Note that the diagram on the far right is particularly simple; the corresponding operator can easily shown to be proportional to $P_0$ using a straightforward generalization of the contraction rules~\eqref{it:gadgetfirst}--\eqref{it:gadgetfourth} we established in Section~\ref{sec:plaquettevertexprod}. Thus reducing the diagram to one of this form is our main technical goal.

Our first standard form $S1$ requires that non-trivial moves (i.e., diagonal arrows) away from the horizontal axis do not mix; that is, the location of the ``active'' arrow may only change from one subdiagram to another when all endpoints are on the horizontal axis. Formally, a diagram has form~$S1$ if there exist indices $\ell_1,\ldots,\ell_p$ such that (setting $\ell_0=0$, $\ell_{p+1}=m$)
\begin{enumerate}\setcounter{enumi}{-1}
\item
All chains of arrows  go through $(\ell_i,0)$, for all $i$.
\item
All non-horizontal arrows 
between $\ell_i$ and $\ell_{i+1}$
are located in a single subdiagram, that is,  there is an $\alpha$ such that $Y^{\alpha}_{\ell}\in \bigcup_j \{W^\alpha_j,(W^\alpha_j)^\dagger\}$  and $Y^{\beta}_{\ell}=\id$ for all $\ell=\ell_{i}+1,\ldots,\ell_{i+1}$ and $\beta\neq\alpha$.
\end{enumerate}
For example, diagram~\eqref{eq:thetaepsY} does not have standard form $S1$, but can be related to the ($S1$-)standard diagram
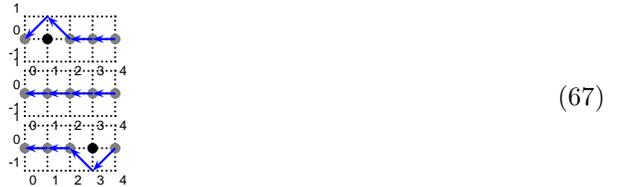
\begin{align}
\begin{matrix}\begin{pspicture}(0,-1)(4,1)
\psgrid[subgriddiv=1,linecolor=lightgray,griddots=5,gridlabels=5pt](0,-1)(4,1)
\rput(0,0){\zzerobig}\rput(4,0){\zzerobig}
\rput(1,0){\zone}\rput(2,0){\zzero}\rput(3,0){\zzero}
\rput(0,0){\zdown}
\rput(3,0){\zleft}
\rput(2,0){\zleft}
\rput(1,1){\zup}
\end{pspicture}\\
\begin{pspicture}(0,-1)(4,1)
\psgrid[subgriddiv=1,linecolor=lightgray,griddots=5,gridlabels=5pt](0,-1)(4,1)
\rput(0,0){\zzerobig}\rput(4,0){\zzerobig}
\rput(1,0){\zzero}\rput(2,0){\zzero}\rput(3,0){\zzero}
\rput(0,0){\zleft}
\rput(1,0){\zleft}
\rput(2,0){\zleft}
\rput(3,0){\zleft}
\end{pspicture}\\
\begin{pspicture}(0,-1)(4,1)
\psgrid[subgriddiv=1,linecolor=lightgray,griddots=5,gridlabels=5pt](0,-1)(4,1)
\rput(0,0){\zzerobig}\rput(4,0){\zzerobig}
\rput(1,0){\zzero}\rput(3,0){\zone}\rput(2,0){\zzero}
\rput(1,0){\zleft}
\rput(2,0){\zup}
\rput(3,-1){\zdown}
\rput(0,0){\zleft}
\end{pspicture}
\end{matrix}\label{eq:thetaepsstandardform}
\end{align}
as the following lemma shows.
\vspace*{12pt}
\noindent
\begin{lemma}\label{lem:firststandardform}
Consider an operator $\Theta(\varepsilon,{\bf Y})\not\in\cup_{\alpha} \{\Theta_n^{\alpha,\uparrow},\Theta_n^{\alpha,\downarrow}\}$ of order $m\leq n$ 
corresponding to a valid diagram $(\varepsilon,{\bf Y})$. Then $\Theta(\varepsilon,{\bf Y})=\Theta(\varepsilon',{\bf Y}')$, where $\Theta(\varepsilon',{\bf Y}')$ is of order $m$ with a valid diagram $(\varepsilon',{\bf Y}')$ of form~$S1$.
\end{lemma}
\vspace*{12pt}
\noindent
{\bf Sketch of Proof:}
This follows by iterative application of Lemma~\ref{lem:substitution}~\eqref{it:firstcomm}. Without loss of generality, assume that the first non-horizontal arrow affects $\cH_{I^{1}}$; e.g., $Y_m^1=(W_0^1)^\dagger$. 
Assume that the horizontal axis in the first subdiagram is reached after application of the operator $Y_\ell^1$ (in our example $Y_\ell^1=W_0^1$.) Clearly, all operators (arrows) between $\ell$ and $m$ affecting  $\cH_{I^{\beta}}$, $\beta\neq \alpha$ can be commuted to the right of $Y_m^1$ (the first upwardpointing arrow in the fist subdiagram) by application of Lemma~\ref{lem:substitution}~\eqref{it:firstcomm}. Recursive application of the same procedure to the resulting diagram (in particular also the part between $0,\ldots,\ell-1$) gives the claim. \square\, 

We will say that a diagram has standard form~$S2$ if it consists of a sequence of ``triangles'' on each subdiagram. That is, the diagram has standard form $S1$ with the additional property
\begin{enumerate}\setcounter{enumi}{1}
\item
$\ell_{i+1}-\ell_{i}=2$ and there is an index $\alpha$ such that $Y^\alpha_{\ell_i+1}Y^\alpha_{\ell_{i+1}}\in \{W^{\alpha}_0(W^{\alpha}_0)^\dagger,(W^{\alpha}_0)^\dagger W^{\alpha}_0\}$.
\end{enumerate}
Note that~\eqref{eq:thetaepsstandardform} already has standard form~$S2$. With Lemma~\ref{lem:firststandardform} and the following result, we can reduce any diagram of interest to a diagram which has standard form~$S2$.
\vspace*{12pt}
\noindent
\begin{lemma}\label{lem:secondstandardform}
Consider an operator $\Theta(\varepsilon,{\bf Y})$ of order $m$ with a valid diagram $(\varepsilon,{\bf Y})$ of form $S1$. Then $\Theta(\varepsilon,{\bf Y})\propto  \Theta(\varepsilon',{\bf Y}')$, where the latter operator is of order $m'\leq m$ and where $(\varepsilon',{\bf Y}')$ is of form $S2$.
\end{lemma}
\vspace*{12pt}
\noindent
{\bf Proof:}
  This follows by applying an appropriate generalization of the contraction rules~\eqref{it:gadgetfirst}--\eqref{it:gadgetthird} discussed in the proof of Theorem~\ref{thm:maingadget}. \square\,

Next we introduce an additional condition: If a diagram has standard form~$S2$ and all non-horizontal arrows corresponding to a subdiagram $\alpha$ are next to each other, we say that the diagram has standard from~$S3$. Formally, this condition can be expressed as
\begin{enumerate}\setcounter{enumi}{2}
\item
There is a sequence of distinct $\alpha_1,\ldots,\alpha_s$ and indices $k_1=0,k_1,\ldots, k_{s-1},k_s=m$ such that
$Y^{\alpha_j}_{k}\in \bigcup_i \{W^{\alpha_j}_i,(W^{\alpha_j}_i)^\dagger\}$ for all $k_j+1\leq k\leq k_{j+1}$.
\end{enumerate}
 Again,~\eqref{eq:thetaepsstandardform} already has standard form~$S3$. We also have the following statement.
\vspace*{12pt}
\noindent
\begin{lemma}\label{lem:thirdstandardform}
Consider an operator
 $\Theta(\varepsilon,{\bf Y})$
of order $m$ corresponding to a valid diagram $(\varepsilon,{\bf Y})$ of form $S2$. Then $\Theta(\varepsilon,{\bf Y})=\Theta(\varepsilon',{\bf Y}')$, where $\Theta(\varepsilon',{\bf Y}')$ is of order $m$ with a valid diagram $(\varepsilon',{\bf Y}')$ of form~$S3$.
\end{lemma}
\vspace*{12pt}
\noindent
{\bf Proof:}
This follows by moving triangles corresponding to different subdiagrams past each other using~\eqref{it:secondcomm} of Lemma~\ref{lem:substitution}. \square\,

We are ready to prove the following generalization of Lemma~\ref{lem:centralproportionality}.
\newtheorem{varlemmaprop}{Lemma~\ref{lem:centralproportionality}$^\prime$}
\vspace*{12pt}
\noindent
\begin{varlemmaprop}
\begin{enumerate}[(a)]
\item\label{it:validoperators}
Let $\Theta(\varepsilon,{\bf Y})\not\in\cup_{\alpha} \{\Theta_n^{\alpha,\uparrow},\Theta_n^{\alpha,\downarrow}\}$ be an operator of order $m\leq n$ corresponding to a valid diagram, where the latter operators are defined by~\eqref{eq:uparrowdiagrams}. Then $\Theta(\varepsilon,{\bf Y})\propto P_0$.
\item\label{it:multicountervalidop}
The $n$-th order operators~\eqref{eq:uparrowdiagrams} are 
\begin{align*}
\Theta_n^{\alpha,\downarrow} &=\Gamma M_0^\alpha\cdots M_{n-1}^\alpha \Gamma\otimes\proj{0}_{\alpha}^{\otimes L}\\
\Theta_n^{\alpha,\uparrow} &=\Gamma (M_{n-1}^\alpha)^\dagger\cdots (M_0^\alpha)^\dagger\Gamma\otimes\proj{0}_{\alpha}^{\otimes L}
\end{align*}
\end{enumerate}
\end{varlemmaprop}
\vspace*{12pt}
\noindent
{\bf Proof:} For the proof of~\eqref{it:validoperators}, we use Lemma~\ref{lem:firststandardform}, Lemma~\ref{lem:secondstandardform} and Lemma~\ref{lem:thirdstandardform}. This allows us to reduce the diagram corresponding to $\Theta(\varepsilon,{\bf Y})$ to standard form~$S3$. But the operator associated to such a diagram is proportional to $P_0$, as can be seen by repeated application of the generalization of rules~\eqref{it:gadgetfirst}--\eqref{it:gadgetfourth} used in the proof of Theorem~\ref{thm:maingadget}.

Statement~\eqref{it:multicountervalidop} follows by inserting the relevant definitions. \square\,

This is all we need to complete the proof of Theorem~\ref{thm:maingadget}$^\prime$. 

\vspace*{12pt}
\noindent{\bf Proof of Theorem~\ref{thm:maingadget}$^\prime$:}
Observe that for every configuration $\varepsilon$ with the property described by~\eqref{eq:uparrowdiagrams}, we have
\begin{align*}
g_m(\varepsilon) &=\sum_{\ell\in\cP_{m-1}}h_\ell(\underbrace{1,\ldots,1}_{n-1})=\sum_{\substack{\ell\in\cP_{m-1}\\
\ell_i\neq 0\textrm{ for all }i}} (-1)^{\sum_i \ell_i}=(-1)^{m-1}
\end{align*}
where we inserted~\eqref{eq:helldef} and the definition~\eqref{eq:cpmdef} of $\cP_{m-1}$. Lemma~\ref{lem:Ausefulform}$^\prime$ and Lemma~\ref{lem:centralproportionality}$^\prime$
 therefore give
\begin{align*} 
\cA^{(m)}&\propto P_0\qquad\textrm{ for } m<n\\
\cA^{(n)}&=const\cdot P_0+ (-1)^{n-1} H_{\textrm{target}}\otimes\proj{0}^{\otimes L}\ .
\end{align*}
The claim of the theorem then follows as in the proof of Theorem~\ref{thm:maingadget}. \square\,


\noindent

\begin{thebibliography}{10}

\bibitem{biamonte08}
J.~D. Biamonte.
\newblock Non-perturbative k-body to two-body commuting conversion hamiltonians
  and embedding problem instances into ising spins.
\newblock {\em Phys. Rev. A}, 77:052331, 2008.

\bibitem{biamontelove08}
J.~D. Biamonte and P.~J. Love.
\newblock Realizable hamiltonians for universal adiabatic quantum computers.
\newblock {\em Phys. Rev. A}, 78:012352, 2008.

\bibitem{bloch58}
C.~Bloch.
\newblock Sur la th\'eorie des perturbations des \'etats li\'es.
\newblock {\em Nuclear Physics}, 6:329--347, 1958.

\bibitem{drinfeld87}
V.G. Drinfeld.
\newblock Quantum groups.
\newblock {\em Proceedings of ICM-86, Berkeley}, 1:798--820, 1987.

\bibitem{duanetal03}
L.-M. Duan, E.~Demler, and M.~D. Lukin.
\newblock Controlling spin exchange interactions of ultracold atoms in optical
  lattices.
\newblock {\em Phys. Rev. Lett.}, 91(9):090402, Aug 2003.

\bibitem{freedmanetal01}
M.~H. Freedman, A.~Kitaev, M.~J. Larsen, and Z.~Wang.
\newblock Topological quantum computation.
\newblock 2001.

\bibitem{jordanfarhi}
S.~P. Jordan and E.~Farhi.
\newblock Perturbative gadgets at arbitrary orders.
\newblock {\em Phys. Rev. A}, 77(062329), 2008.

\bibitem{kempekitaevregev06}
J.~Kempe, A.~Kitaev, and O.~Regev.
\newblock The complexity of the local hamiltonian problem.
\newblock {\em SIAM Journal of Computing}, 35:1070, 2006.

\bibitem{kitaev06}
A.~Kitaev.
\newblock Anyons in an exactly solved model and beyond.
\newblock {\em Annals of Physics}, 321(2), 2006.

\bibitem{kitaev97}
A.~Y. Kitaev.
\newblock Fault-tolerant quantum computation by anyons.
\newblock {\em Ann. Phys.}, (303):2--30, 2003.

\bibitem{kitaevetalbook02}
A.~Y. Kitaev, A.~H. Shen, and M.~N. Vyalvi.
\newblock {\em Classical and Quantum Computation}, volume~47 of {\em Graduate
  Studies in Mathematics}.
\newblock American Mathematical Society, July 2002.

\bibitem{LevinWen04}
X.-G.~Wen M.~A.~Levin.
\newblock String-net condensation: A physical mechanism for topological phases.
\newblock {\em Phys. Rev.}, B71(045110), 2005.

\bibitem{michelietal06}
A.~Micheli, G.~K. Brennen, and P.~Zoller.
\newblock A toolbox for lattice-spin models with polar molecules.
\newblock {\em Nature Physics}, 2:341--347, 2006.

\bibitem{mochon03}
C.~Mochon.
\newblock Anyons from non-solvable finite groups are sufficient for universal
  quantum computation.
\newblock {\em Phys. Rev. A}, 67(022315), 2003.

\bibitem{mochon04}
C.~Mochon.
\newblock Anyon computers with smaller groups.
\newblock {\em Phys. Rev. A}, 69(032306), 2004.

\bibitem{ogburnpreskill99}
R.~W. Ogburn and J.~Preskill.
\newblock Topological quantum computation.
\newblock {\em LLNCS}, 1509:341--356, 1999.

\bibitem{oliveiraterhal05}
R.~Oliveira and B.~M. Terhal.
\newblock The complexity of quantum spin systems on a two-dimensional square
  lattice.
\newblock {\em Quantum Information and Computation}, 8(10):900--924, 2008.

\bibitem{preskill97}
J.~Preskill.
\newblock Fault-tolerant quantum computation.
\newblock {\em Introduction to Quantum Computation}, edited by H.-K.~Lo, S.~Popescu, and T.~P.~Spiller, arXiv.org:quant-ph/9712048v1,  1998.

\end{thebibliography}
\end{document}